\documentclass[3p]{elsarticle}

\usepackage{lineno,hyperref}
\usepackage[T1]{fontenc}
\usepackage[utf8]{inputenc}
\usepackage{lmodern}
\usepackage[miktex]{gnuplottex}
\usepackage{graphicx}
\usepackage{amsmath}
\usepackage{bm}
\usepackage{algorithm}
\usepackage{listings}
\usepackage{algpseudocode}
\usepackage{xfrac}
\usepackage{amssymb}
\usepackage{mathtools}

\usepackage{color}

\newcommand{\refFig}[1] {Fig.~\ref{#1}}
\newcommand{\refTab}[1] {Table \ref{#1}}
\newcommand{\refEqu}[1] {Eq.~(\ref{#1})}

 \newcommand{\bfE}{\mathbf{E}}
\newcommand{\bE}{\mathbf{E}}
\newcommand{\bfB}{\mathbf{B}}
 
\newcommand{\bB}{\mathbf{B}}

\newcommand{\bJ}{\mathbf{J}}
\newcommand{\bu}{\mathbf{u}}

\newcommand{\bv}{\mathbf{v}}

\newcommand{\br}{\mathbf{r}}
\newcommand{\bs}{\mathbf{s}}
\newcommand{\dd}{\rm{d}}

\modulolinenumbers[5]

\journal{Computer Physics Communications}

\begin{document}

\begin{frontmatter}

\title{ECsim-CYL: Energy Conserving Semi-Implicit particle in cell simulation in   axially symmetric  cylindrical coordinates}

\author[inst1]{Diego Gonzalez-Herrero}
\author[inst1]{Alfredo Micera}
\author[inst1]{Elisabetta Boella}
\author[inst2]{Jaeyoung Park}
\author[inst1]{Giovanni Lapenta\corref{mycorrespondingauthor}}
\cortext[mycorrespondingauthor]{Corresponding author}
\ead{giovanni.lapenta@kuleuven.be}

\address[inst1]{Department of Mathematics, KU Leuven, University of Leuven, Belgium}
\address[inst2]{Energy Matter Conversion Corporation (EMC2), San Diego, California (USA)}

\begin{abstract}

Based on the previously developed Energy Conserving Semi Implicit Method (ECsim) code, we present 
its cylindrical implementation, called ECsim-CYL, to be used for axially symmetric problems.  
The main motivation for the development of the cylindrical version is to greatly improve the computational 
speed  by utilizing cylindrical symmetry.  The ECsim-CYL discretizes the field equations in two-dimensional cylindrical coordinates using the finite volume method .  For the particle mover, it uses a modification of ECsim's mover for cylindrical coordinates by keeping track of all three components of velocity vectors, while only keeping radial and axial coordinates of particle positions.  In this paper, we describe the details 
of the algorithm used in the ECsim-CYL and present  a series of tests to validate the accuracy of the code 
 including a wave spectrum in a homogeneous plasmas inside a cylindrical waveguide and free 
expansion of a spherical plasma ball in vacuum. The ECsim-CYL retains the stability properties of ECsim 
and conserves the energy within  machine precision, while accurately describing the plasma behavior 
in the test cases.

\end{abstract}


\end{frontmatter}

\section{Introduction}
Particle In Cell (PIC) algorithms are widely used to study natural and laboratory plasmas using fully kinetic description \citep{birdsall-langdon,hockney-eastwood}.
In a typical PIC algorithm, particles and fields are advanced alternatively in a cycle. 
Particles are moved with the fields obtained by interpolation from a grid 
and the particle information is then used to project the sources of the fields, i.e. current and charge, onto the grid advancing the field values in time. 
This alternation of particle movement and field solution is the typical feature of explicit PIC codes.  
Explicit PIC codes are straight-forward to design and lead to extremely efficient parallel implementations.  
As such, explicit PIC codes are always among the most successful applications on new supercomputers. 

However, explicit PIC codes suffer from a scale limitation \citep{birdsall-langdon,hockney-eastwood}: 
the alternative field and particle solver approach of the explicit cycle leads to two limitations: 
the time step needs to resolve the electron plasma frequency and the grid spacing need to resolve a multiple of the Debye length (the multiple depends on the interpolations scheme).  
Plasma problems are typically multi scale, with ions responding on much longer and slower scales than electrons.  
The need for explicit PIC codes to resolve all electron scales, led to development of implicit PIC methods to overcome this difficulty.

Implicit methods retain the coupling of particle motion and field solution at each time step. 
They are  no longer done separately in sequence but rather become part of an iteration. 
This coupling is non-linear and it requires the use of non-linear solvers \citep{markidis2011energy,Chen-jcp-2011}. 

However, when the coupling of particles and fields is linearized, the iteration becomes linear and typically based on Krylov solvers: 
in this case the methods are considered semi-implicit. This semi-implicitness should not be confused with the methods that solve 
implicitly only the field equations but still remain explicit in alternating particle mover and field solver \citep{morse1971numerical}. 
This class of algorithms are designed to overcome the speed of light limitation in the electromagnetic waves and are not considered here. 

All implicit and semi-implicit PIC codes enable the user to use larger time steps and grid spacing, an advantage when dealing with multi scale problems where the electron physics need not be represented in detail but still needs to be captured in macroscopic terms \citep{lapenta2012particle}. For these algorithms, the electron response to large scales is captured correctly but the small scale fluctuations (e.g. Langmuir waves) are damped \citep{brackbill-forslund,Brackbill:1985}. While fully implicit methods require an advanced arsenal of non-linear solvers, supported by efficient preconditioning \citep{chen2014fluid} to reduce the number of iterations, semi-implicit methods can use more standard linear solvers. 

Semi-implicit methods were the first to be deployed in large practical applications. The literature reports two lines of semi-implicit codes: the direct implicit method \citep{directimplicit} and the moment implicit methods \citep{brackbill-forslund}. The most modern implementations are, respectively, LSP \citep{welch2004implementation} and iPic3D \citep{ipic3d}.  Among the semi-implicit algorithms, we have recently developed a new algorithm called ECsim \citep{lapenta2016exactly,Lapenta2017,GonzalezHerrero2018} that adds one more desirable property to the previous semi-implicit schemes: energy conservation.

ECsim  uses a new tool to represent the linear response of the plasma to the fields: the mass matrix \citep{burgess1992mass,lapenta2016exactly}. The use of a new mover allows the introduction of this linear formulation without requiring any linearization: the result is that ECsim is exactly energy conserving. This has two aspects. First, there is a formal analytical energy integral that has been derived to show energy conservation by the scheme. Second, when coded into a computer implementation, energy is conserved to machine precision, a powerful debugging feature as well as a desirable physical property \citep{lapenta2016exactly}. 

ECsim has the extra computational complexity of requiring the computation of the mass matrix by interpolation from the particles. The extra cost of this operation was evaluated in Ref. \citep{GonzalezHerrero2018}. Compared with other PIC codes, ECsim requires more memory to store the mass matrix and more computational costs to compute it. When compared with specifically with the  moment implicit methods, such as iPic3D \citep{ipic3d}, ECsim increases the number of operations needed to project particle information to the grid, as required to compute the mass matrix. However, this increase is in part compensated by the reduced complexity of the mover. The cost per cycle increases by a factor of 2 to 4 but it is more than compensated by the ability to use larger time steps allowed by energy conservation \cite{GonzalezHerrero2018}. The increase in memory is not significant as it does not increase the particle memory, by far the overwhelming majority of the memory used: by comparison, the additional memory required to store the mass matrix is not significant. 

In many applications, there is an easy way to reduce the cost: use the intrinsic symmetry of the problem. Many experiments and natural processes have a natural axis of symmetry and cylindrical symmetry allows one to treat them as two-dimensional problems.
Symmetry reduces the computational effort by reducing the dimensionality, enabling  higher spatial resolutions or reducing the computational cost.
We have implemented a new version of ECsim in cylindrical coordinates (hereinafter ECsim-CYL) for these axially
symmetric cases. This new version, analogously to the original ECsim code, conserves
the total energy exactly and retains the same stability properties.

Cylindrical symmetry has been used in several previous explicit and even semi-implicit codes, based on simpler algorithms that do not conserve energy exactly. In the explicit 
approach, cylindrical symmetry is described in textbooks \citep{birdsall-langdon} and has been applied in production codes \citep{Ringle2011}. Also for the moment implicit approach a cylindrical version was developed \citep{Wallace1986}, based on the same formulation of the implicit moment method \citep{Brackbill:1985} in cartesian geometry, at the basis of Venus, Celeste and iPic3D: this method does not conserve energy exactly. 

However, ECsim-CYL is the first semi-implicit code which implements the Energy Conserving Semi
Implicit algorithm in cylindrical coordinates and is the first semi-implicit code that   conserves energy. 

More recently, general curvilinear geometry codes have been developed both for explicit  \citep{lapenta2011democritus,Delzanno2013} and for implicit 
codes \citep{Chacon2016}.

ECsim-CYL has been developed in C++ and uses MPI for parallel communications. It shares the infrastructure
(particle communication, particle-to-grid interpolations, etc.) of the fully 3D Cartesian ECsim \citep{GonzalezHerrero2018} and 
 has a similar scaling behavior and performance. 
 
The paper is organized as follows. In section 2, we describe the algorithm used in the ECsim-CYL implementation regarding 
the discretization of the field equations using cylindrical coordinates and the implementation of the particle mover. 
In section 3, we test the accuracy of both the field solver and particle mover.  In section 4, we test the ECsim-CYL results 
against the analytical formula regarding the plasma wave in an homogeneous plasma inside a cylindrical waveguide.  This is followed 
by the test results of ECsim-CYL in the simulation of homogeneous expansion of a plasma sphere in vacuum with different spatio-temporal resolution 
and ion mass ratio in section 5.  In Section 6, we summarize the results and present the conclusions.

 \section{Description of the algorithm}

The scheme and the main loop of the code are identical to the ones of ECsim described 
in \cite{lapenta2016exactly, GonzalezHerrero2018}. The interpolation of the current densities from
the particles to the grid and the calculation of the mass matrices are performed in the same way
as in Cartesian coordinates. The only thing to keep in mind is that now the volume of the
individual cells is not constant in space, but depends on the radial position of the cell. 

The particle initialization is, however, performed in a slightly different manner. In order to have an 
homogeneous density (if that is what's desired), the charge in each cell has to be proportional to its volume. 
This can be achieved using a different number of particles per cell or assigning a weight to the particles 
depending on which cell they belong to. Let's assume we want to have a charge density $\rho$. In the first 
case, we define the number of particles in the first cell $n_0^{\text{cell}}$ and then the charge of each 
particle is given by: $q_p = \rho \, V_{0} / n_0^{\text{cell}}$, where $V_{0}$ is the volume of the first cell 
(at $r=0$). Then the number of particles in the $i^{\text{th}}$ cell (with volume $V_i$) is computed as 
$n_i^{\text{cell}} = n_0^{\text{cell}} \, V_{i}/V_{0}$. In the second case, the number of particles 
$n^{\text{cell}}$ is the same for all cells, and the charge of the particles depends on which cell they are 
at the beginning of the simulation: all particles initially in the $i^\text{th}$ cell will have a charge according to
 $q_p = \rho \, V_{i} / n^{\text{cell}}$.

In order to have a significant number of particles in all the cells, more particles are needed overall in 
the version of the algorithm in which all the individual particles are assigned the same weight. Both options 
are available in ECsim-CYL and both of them have given similar results in the test cases. 

The main differences between ECsim-CYL and the Cartesian version are to be found in the particle mover and 
in the discretization of the field equations, which are described in detail in the next subsections.

\subsection{Field solver}
\label{sec.2.1}
The field equations are the same as in the Cartesian version, however the discretization technique
used is different. While in ECsim we used the Finite Difference method, here we have employed a Finite 
Volumes technique to integrate the equations given by Eqns.~(\ref{eq-Bfield}) and (\ref{eq-Efield}):
\begin{eqnarray}
  \label{eq-Bfield}
  && \nabla \times \bfE^{n+\theta} + \frac{1}{c} \frac{\bfB^{n+\theta}-\bfB^n}{\theta \Delta t} =0,\\
  \label{eq-Efield}
  && \nabla \times \bfB^{n+\theta} - \frac{1}{c} \frac{\bfE^{n+\theta}-\bfE^n}{\theta \Delta t} =\frac{4\pi}{c} \tilde{\bJ},
\end{eqnarray}
where $\bfE$ is the electric field (defined on the nodes), $\bfB$ the magnetic field (defined on the cell centers) and $\theta$ 
is the decentering parameter (which has to be 0.5 to conserve the energy). $\tilde{\bJ}$ is the implicit current (defined on the
nodes) which is computed
through the mass matrices: $\tilde{\bJ} = \widehat{\bJ}^{n}+\sum_{N^\prime} M_{NN^\prime} \bE^{n+\theta}$. A detailed description can be
found in \cite{lapenta2016exactly}.

We have already stated above that to discretize Eqns.~(\ref{eq-Bfield}) and (\ref{eq-Efield}) we use the Finite-Volume method \cite{leveque2}. 
We integrate each term over the control volume, i.e., the volume of the cell ($V_c$) for \refEqu{eq-Bfield} and the volume of the node ($V_n$) for \refEqu{eq-Efield}.
Let's denote with (i,j) the indices of the cell/node in the r and z direction. Then the control volumes are, respectively
\begin{eqnarray}
  V_c(i_c,j_c) &=& \pi\,\Delta z\,(\Delta r)^2\,((i_c+1)^2 - i_c^2) ,\\
  V_n(i_n,j_n) &=& \pi\,\Delta z\,(\Delta r)^2\,((i_n-1/2)^2 - (i_n+1/2)^2)
\end{eqnarray}

Integrating \refEqu{eq-Bfield} and \refEqu{eq-Efield} over the corresponding control volume we have
\begin{eqnarray}
  && \int_{V_c} \nabla \times \bfE^{n+\theta} dV+ V_c \; \frac{1}{c} \frac{\bfB^{n+\theta}-\bfB^n}{\theta \Delta t} =0,\\
  && \int_{V_n} \nabla \times \bfB^{n+\theta} dV- V_n \; \frac{1}{c} \frac{\bfE^{n+\theta}-\bfE^n}{\theta \Delta t} = V_n \; \frac{4\pi}{c} \tilde{\bJ},
\end{eqnarray}
where we have made use of the result that, for any magnitude $A$ defined on the nodes, the integral over  $V_n$ is 
\begin{equation}
  \int_{V_{n}} A_{n} dV= V_n \; A_n,
\end{equation}
and the same for the magnitudes defined on the centers and the integral over $V_c$.

To compute the integral of $\nabla \times \bfE^{n+\theta}$ we can use the fact that 
\begin{equation}
  \int_{V_{c}} \nabla \times \bfE^{n+\theta} dV= \int_{S} \hat{\bf n} \times \bfE^{n+\theta} dS
\end{equation}

\begin{figure}[h]
  \center
  \includegraphics[width=2.0in]{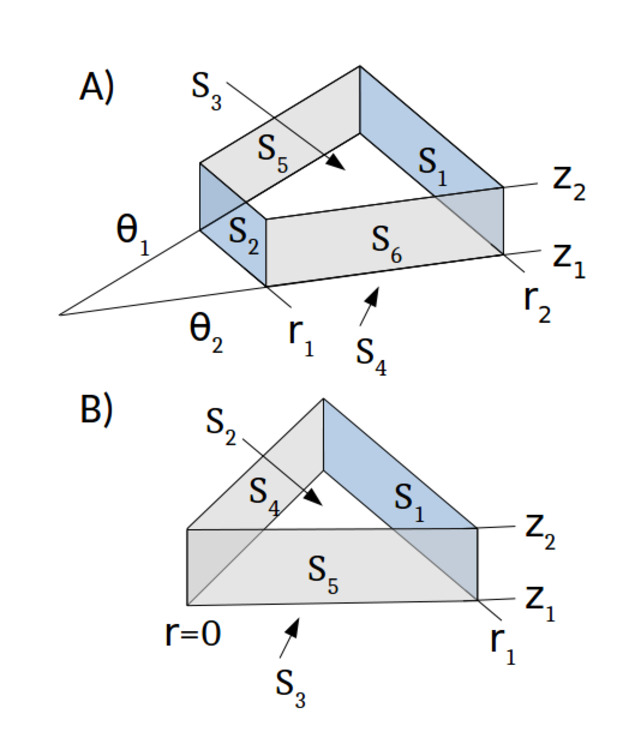}
  \caption{Scheme of the volume discretization used in the field solver for a generic cell (A) and for the
          first cell in the r direction (B).}
  \label{fig-cell}
\end{figure}

In the following we will omit the super index $\{\cdot\}^{n+\theta}$ for simplicity. Using the scheme 
shown in \refFig{fig-cell} (A) we can expand the integral of the curl as follows
\begin{eqnarray*}
\int_{V_{c}} \nabla \times \bfE &dV =& \int_{S_1} \hat{\bf r} \times \bE dS- \int_{S_2} \hat{\bf r} \times \bE dS\\
                                &+& \int_{S_3} \hat{\bf z} \times \bE dS- \int_{S_4} \hat{\bf z} \times \bE dS\\
                                &+& \int_{S_5} \hat{\boldsymbol{\phi}} \times \bE dS- \int_{S_6} \hat{\bf \phi} \times \bE dS,
\end{eqnarray*}
where $\hat{\bf r}$, $\hat{\boldsymbol{\phi}}$ and $\hat{\bf z}$ are the unitary vectors in the $r$, $z$ and $\phi$ directions. After
integrating over each surface we arrive at the following expressions:
\begin{eqnarray}
  \left(\int_{V_{c}} \nabla \times \bfE dV\right)_r   &=& 2 \pi [I_r(E_\phi(r, z_2)) - I_r(E_\phi(r, z_1))],\\
  \left(\int_{V_{c}} \nabla \times \bfE dV \right)_\phi &=& 2 \pi [I_z(r_1,E_z(r_1, z)) + I_z(r_2,E_z(r_2, z)) \nonumber \\
                                                        && \qquad - I_r(E_r(r, z_2)) - I_r(E_r(r, z_1))],\\
  \left(\int_{V_{c}} \nabla \times \bfE dV \right)_z    &=& 2 \pi [I_z(r_2,E_\phi(r_2, z)) - I_z(r_1,E_\phi(r_1, z))],
\end{eqnarray}
where 
\begin{eqnarray}
  I_r(A(r,z)) &=& \int_{r_1}^{r_2} r A(r,z) \dd r \nonumber \\
              &=& A(r_1,z) \frac{\bar{r}^2 - r_1^2}{2} + A(r_2,z) \frac{\bar{r}^2 - r_2^2}{2} ,\\
  I_z(r,A(r,z)) &=& r \int_{z_1}^{z_2} A(r,z) \dd z \nonumber \\
                &=& \frac{\Delta z}{2}(A(r,z_1) + A(r,z_2)),
\end{eqnarray}
with $\bar{r} = (r_1 + r_2)/2$.
The same can be applied to the curl of $\bB$ apart from the case with r=0, which has to be treated separately. 
Using the scheme shown in  \refFig{fig-cell} (B) and following the same procedure we can obtain:
\begin{eqnarray}
  \left(\int_{V_{c}} \nabla \times \bfB dV \right)_r^{r=0}    &=& 0, \\
  \left(\int_{V_{c}} \nabla \times \bfB dV \right)_\phi^{r=0} &=& 0, \\
  \left(\int_{V_{c}} \nabla \times \bfB dV \right)_z^{r=0}    &=& 2 \pi I_z(r_1,B_\phi(r_1, z))],
\end{eqnarray}

With this procedure, $\nabla \times \bfE$ (which is defined on the cell centers) can be computed from the values of $\bE$ on the nodes. 
And similarly, $\nabla \times \bfB$ is defined on the nodes and is computed from the values of $\bB$ on the cell centers. This way,
every term in \refEqu{eq-Efield} is defined on the nodes and the ones in \refEqu{eq-Bfield} is defined on the cell centers. In addition,
this discretization ensures the conservation of the total energy.

To solve Eqns.~(\ref{eq-Bfield}) and (\ref{eq-Efield}) we use the GMRES method implemented in PETSc \citep{petsc-web-page}.

\subsection{Particle mover}
\label{sec.2.2}
We employ a combination of the ECsim mover described in \citep{lapenta2016exactly} with the Boris mover in 
cylindrical coordinates \citep{Wallace1986}. Once the electric and magnetic fields at time $n+\theta$ have
been computed, we advance the particles as in Cartesian coordinates. First, we build the position and 
velocity of the particles in Cartesian coordinates using the values in cylindrical coordinates: $\bu_p^{n} = (v_r^{n}, v_\phi^{n}, v_z^{n})$ and 
$\bs_p^{n+1/2} = (r^{n+1/2}, 0, z^{n+1/2})$.  And then we operate as in the Cartesian case:
\begin{eqnarray}
    \bar{\bu}_p &=& \alpha^n \left(\bu_p^n + \beta_s \bfE_p^{n+\theta}(\bs_p^{n+1/2})\right), \\
    \bu_p^{n+1} &=& 2 \bar{\bu_p} - \bu_p^n, \\
    \bs_p^{n+3/2} &=& \bs_p^{n+1/2} + \Delta t \bu_p^{n+1},
\end{eqnarray}
where $\bfE_p^{n+\theta}$, $\bfB_p^{n+\theta}$ are the electric and magnetic fields at time $n+\theta$ at the
particle position. $\Delta t$ is the time step and $\alpha$ is the rotation matrix given by:
\begin{equation}
    {\alpha}_p^n =  \frac{1}{1+(\beta_s B_p^{n})^2} \left(\mathbb{I}-\beta_s \mathbb{I} \times \bB_p^n +\beta_s^2
                    \bB_p^n \bB_p^n \right),
\end{equation}
where $\beta=q_p = \Delta t/(2 m_p)$, with $m_p$ and $q_p$ the mass and charge of the particle. After this 
we have the new position and velocity : $\bs_p^{n+3/2} = (x^{n+3/2}, y^{n+3/2}, z^{n+3/2})$ and 
$\bu_p^{n+1} = (u_x^{n+1}, u_y^{n+1}, u_z^{n+1})$.

Then, we rotate the particle back to the plane $y = 0$, as is done in the Boris mover \citep{Wallace1986}.
\begin{eqnarray}
  \alpha &=& \arctan\left(\frac{u_y^{n+1}}{u_x^{n+1}}\right), \\
  R_z(\alpha) &=& \left(\begin{matrix}
                  \cos(\alpha) & -\sin{\alpha} & 0 \\
                  \sin(\alpha) &  \cos{\alpha} & 0 \\
                  0            &  0            & 1
                \end{matrix} \right).
\end{eqnarray}
The velocity and position in cylindrical coordinates are obtained applying this rotation to the vectors in
Cartesian coordinates:
\begin{eqnarray}
  \label{eq_mover_v}
  \bv_p^{n+1}   &=& R_z(\alpha) \; \bu_p^{n+1}, \\
  \label{eq_mover_r}
  \br_p^{n+3/2} &=& R_z(\alpha) \; \bs_p^{n+3/2} .
\end{eqnarray}

Note that by the end of the mover step the azimuthal component of the position is zero, so the particles are
always in a plane. This procedure conserves the energy and does not spoil the stability features of the
original algorithm in Cartesian coordinates.

 \section{Accuracy of the field solver and particle mover}
A very important practical concept is the order of accuracy of each component of the algorithm. The order of accuracy has obviously a practical impact on the fidelity of the solution but in a particle in cell code this aspect is complicated by the noise produced by the particles. One should really consider ensemble of simulations and discuss convergence in statistical terms. However, for practical purposes the order of accuracy is a powerful tool for the verification of a code. For ECsim, energy conservation is itself an extremely useful debugging tool. As the code was being developed, we experimented how sensitive energy conservation is to any bug in the code, from communication issues to more traditional coding errors. 

Order of accuracy can be an additional tool to identify implementation errors in each component of the code \citep{riva2017methodology}. In order to determine the accuracy of the cylindrical implementation, code results are compared with reference solutions, 
either from the Matlab function ODE45 or from the analytical calculations. Detailed parameter scans have been conducted 
by varying time steps and computing the errors with respect to the reference solution. Below we show that both particle mover and  field 
solver have  2nd order accuracy.  

\subsection{Particle mover}
The trajectory of a particle with mass $m$ and charge $q$ in a non uniform radial electric field and axial 
magnetic field is investigated. We assume that both $E_r$ and $B_z$ are linear with the radial coordinate $r$:
\begin{eqnarray}
  E_r=E_{r0} + k \, r \\
  B_z=B_{z0} + h \, r
\end{eqnarray}
where $E_{r0}$ and $B_{z0}$ are the values of the fields at $r=0$ and $h$ and $k$ are constants which determine the linear
dependence of the fields. The rest of the components of the electric and magnetic fields are assumed to be zero.

\begin{figure}[h]
  \center
  \begingroup
  \makeatletter
  \providecommand\color[2][]{    \GenericError{(gnuplot) \space\space\space\@spaces}{      Package color not loaded in conjunction with
      terminal option `colourtext'    }{See the gnuplot documentation for explanation.    }{Either use 'blacktext' in gnuplot or load the package
      color.sty in LaTeX.}    \renewcommand\color[2][]{}  }  \providecommand\includegraphics[2][]{    \GenericError{(gnuplot) \space\space\space\@spaces}{      Package graphicx or graphics not loaded    }{See the gnuplot documentation for explanation.    }{The gnuplot epslatex terminal needs graphicx.sty or graphics.sty.}    \renewcommand\includegraphics[2][]{}  }  \providecommand\rotatebox[2]{#2}  \@ifundefined{ifGPcolor}{    \newif\ifGPcolor
    \GPcolortrue
  }{}  \@ifundefined{ifGPblacktext}{    \newif\ifGPblacktext
    \GPblacktextfalse
  }{}  \let\gplgaddtomacro\g@addto@macro
  \gdef\gplbacktext{}  \gdef\gplfronttext{}  \makeatother
  \ifGPblacktext
    \def\colorrgb#1{}    \def\colorgray#1{}  \else
    \ifGPcolor
      \def\colorrgb#1{\color[rgb]{#1}}      \def\colorgray#1{\color[gray]{#1}}      \expandafter\def\csname LTw\endcsname{\color{white}}      \expandafter\def\csname LTb\endcsname{\color{black}}      \expandafter\def\csname LTa\endcsname{\color{black}}      \expandafter\def\csname LT0\endcsname{\color[rgb]{1,0,0}}      \expandafter\def\csname LT1\endcsname{\color[rgb]{0,1,0}}      \expandafter\def\csname LT2\endcsname{\color[rgb]{0,0,1}}      \expandafter\def\csname LT3\endcsname{\color[rgb]{1,0,1}}      \expandafter\def\csname LT4\endcsname{\color[rgb]{0,1,1}}      \expandafter\def\csname LT5\endcsname{\color[rgb]{1,1,0}}      \expandafter\def\csname LT6\endcsname{\color[rgb]{0,0,0}}      \expandafter\def\csname LT7\endcsname{\color[rgb]{1,0.3,0}}      \expandafter\def\csname LT8\endcsname{\color[rgb]{0.5,0.5,0.5}}    \else
      \def\colorrgb#1{\color{black}}      \def\colorgray#1{\color[gray]{#1}}      \expandafter\def\csname LTw\endcsname{\color{white}}      \expandafter\def\csname LTb\endcsname{\color{black}}      \expandafter\def\csname LTa\endcsname{\color{black}}      \expandafter\def\csname LT0\endcsname{\color{black}}      \expandafter\def\csname LT1\endcsname{\color{black}}      \expandafter\def\csname LT2\endcsname{\color{black}}      \expandafter\def\csname LT3\endcsname{\color{black}}      \expandafter\def\csname LT4\endcsname{\color{black}}      \expandafter\def\csname LT5\endcsname{\color{black}}      \expandafter\def\csname LT6\endcsname{\color{black}}      \expandafter\def\csname LT7\endcsname{\color{black}}      \expandafter\def\csname LT8\endcsname{\color{black}}    \fi
  \fi
    \setlength{\unitlength}{0.0500bp}    \ifx\gptboxheight\undefined      \newlength{\gptboxheight}      \newlength{\gptboxwidth}      \newsavebox{\gptboxtext}    \fi    \setlength{\fboxrule}{0.5pt}    \setlength{\fboxsep}{1pt}\begin{picture}(4818.00,2266.00)    \gplgaddtomacro\gplbacktext{      \csname LTb\endcsname      \put(688,512){\makebox(0,0)[r]{\strut{}$0$}}      \put(688,824){\makebox(0,0)[r]{\strut{}$0.05$}}      \put(688,1136){\makebox(0,0)[r]{\strut{}$0.1$}}      \put(688,1449){\makebox(0,0)[r]{\strut{}$0.15$}}      \put(688,1761){\makebox(0,0)[r]{\strut{}$0.2$}}      \put(688,2073){\makebox(0,0)[r]{\strut{}$0.25$}}      \put(784,352){\makebox(0,0){\strut{}$0$}}      \put(1533,352){\makebox(0,0){\strut{}$2$}}      \put(2282,352){\makebox(0,0){\strut{}$4$}}      \put(3031,352){\makebox(0,0){\strut{}$6$}}      \put(3780,352){\makebox(0,0){\strut{}$8$}}      \put(4529,352){\makebox(0,0){\strut{}$10$}}    }    \gplgaddtomacro\gplfronttext{      \csname LTb\endcsname      \put(128,1292){\rotatebox{-270}{\makebox(0,0){\strut{}r ($d_i$)}}}      \put(2656,112){\makebox(0,0){\strut{}t ($\omega_{pi}^{-1}$)}}      \csname LTb\endcsname      \put(2205,1056){\makebox(0,0)[r]{\strut{}$\Delta$t = 0.1}}      \csname LTb\endcsname      \put(2205,896){\makebox(0,0)[r]{\strut{}$\Delta$t = 0.001}}      \csname LTb\endcsname      \put(2205,736){\makebox(0,0)[r]{\strut{}Exact Sol.}}    }    \gplbacktext
    \put(0,0){\includegraphics{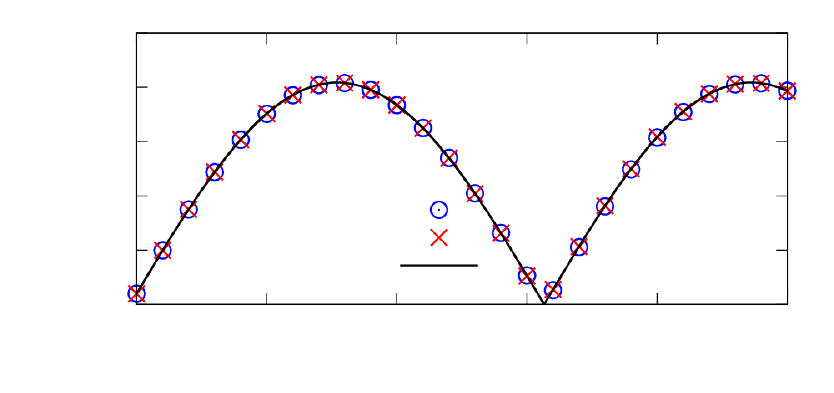}}    \gplfronttext
  \end{picture}\endgroup
   \caption[Evolution of the radial position]{Evolution of the radial position of a particle with $m=1$, $q=1$, 
           $\mathbf{r}=(r,\theta, z)=(0.01,0,0)$ and $\mathbf{v}=(v_r,v_{theta}, v_z)=(0.1,0,0)$ under the influence
           of an electromagnetic field characterized by $E_{r0}=0.01$, $B_{z0}=1$, $k=0.01$ and $h=1$. All quantities 
           are expressed in arbitrary units. The black solid line indicates results obtained with {\tt ODE45}, while the 
           red and blue lines indicate results obtained with the ECsim-CYL particle mover for $\Delta t=0.001$ and $\Delta t=0.1$, respectively.
           The error for different $\Delta t$ is shown in \refFig{figu.9}}
  \label{figu.7}
\end{figure}

\begin{figure}
  \center
  \begingroup
  \makeatletter
  \providecommand\color[2][]{    \GenericError{(gnuplot) \space\space\space\@spaces}{      Package color not loaded in conjunction with
      terminal option `colourtext'    }{See the gnuplot documentation for explanation.    }{Either use 'blacktext' in gnuplot or load the package
      color.sty in LaTeX.}    \renewcommand\color[2][]{}  }  \providecommand\includegraphics[2][]{    \GenericError{(gnuplot) \space\space\space\@spaces}{      Package graphicx or graphics not loaded    }{See the gnuplot documentation for explanation.    }{The gnuplot epslatex terminal needs graphicx.sty or graphics.sty.}    \renewcommand\includegraphics[2][]{}  }  \providecommand\rotatebox[2]{#2}  \@ifundefined{ifGPcolor}{    \newif\ifGPcolor
    \GPcolortrue
  }{}  \@ifundefined{ifGPblacktext}{    \newif\ifGPblacktext
    \GPblacktextfalse
  }{}  \let\gplgaddtomacro\g@addto@macro
  \gdef\gplbacktext{}  \gdef\gplfronttext{}  \makeatother
  \ifGPblacktext
    \def\colorrgb#1{}    \def\colorgray#1{}  \else
    \ifGPcolor
      \def\colorrgb#1{\color[rgb]{#1}}      \def\colorgray#1{\color[gray]{#1}}      \expandafter\def\csname LTw\endcsname{\color{white}}      \expandafter\def\csname LTb\endcsname{\color{black}}      \expandafter\def\csname LTa\endcsname{\color{black}}      \expandafter\def\csname LT0\endcsname{\color[rgb]{1,0,0}}      \expandafter\def\csname LT1\endcsname{\color[rgb]{0,1,0}}      \expandafter\def\csname LT2\endcsname{\color[rgb]{0,0,1}}      \expandafter\def\csname LT3\endcsname{\color[rgb]{1,0,1}}      \expandafter\def\csname LT4\endcsname{\color[rgb]{0,1,1}}      \expandafter\def\csname LT5\endcsname{\color[rgb]{1,1,0}}      \expandafter\def\csname LT6\endcsname{\color[rgb]{0,0,0}}      \expandafter\def\csname LT7\endcsname{\color[rgb]{1,0.3,0}}      \expandafter\def\csname LT8\endcsname{\color[rgb]{0.5,0.5,0.5}}    \else
      \def\colorrgb#1{\color{black}}      \def\colorgray#1{\color[gray]{#1}}      \expandafter\def\csname LTw\endcsname{\color{white}}      \expandafter\def\csname LTb\endcsname{\color{black}}      \expandafter\def\csname LTa\endcsname{\color{black}}      \expandafter\def\csname LT0\endcsname{\color{black}}      \expandafter\def\csname LT1\endcsname{\color{black}}      \expandafter\def\csname LT2\endcsname{\color{black}}      \expandafter\def\csname LT3\endcsname{\color{black}}      \expandafter\def\csname LT4\endcsname{\color{black}}      \expandafter\def\csname LT5\endcsname{\color{black}}      \expandafter\def\csname LT6\endcsname{\color{black}}      \expandafter\def\csname LT7\endcsname{\color{black}}      \expandafter\def\csname LT8\endcsname{\color{black}}    \fi
  \fi
    \setlength{\unitlength}{0.0500bp}    \ifx\gptboxheight\undefined      \newlength{\gptboxheight}      \newlength{\gptboxwidth}      \newsavebox{\gptboxtext}    \fi    \setlength{\fboxrule}{0.5pt}    \setlength{\fboxsep}{1pt}\begin{picture}(4818.00,2266.00)    \gplgaddtomacro\gplbacktext{      \csname LTb\endcsname      \put(784,512){\makebox(0,0)[r]{\strut{}$-0.15$}}      \put(784,772){\makebox(0,0)[r]{\strut{}$-0.1$}}      \put(784,1032){\makebox(0,0)[r]{\strut{}$-0.05$}}      \put(784,1293){\makebox(0,0)[r]{\strut{}$0$}}      \put(784,1553){\makebox(0,0)[r]{\strut{}$0.05$}}      \put(784,1813){\makebox(0,0)[r]{\strut{}$0.1$}}      \put(784,2073){\makebox(0,0)[r]{\strut{}$0.15$}}      \put(880,352){\makebox(0,0){\strut{}$0$}}      \put(1610,352){\makebox(0,0){\strut{}$2$}}      \put(2340,352){\makebox(0,0){\strut{}$4$}}      \put(3069,352){\makebox(0,0){\strut{}$6$}}      \put(3799,352){\makebox(0,0){\strut{}$8$}}      \put(4529,352){\makebox(0,0){\strut{}$10$}}    }    \gplgaddtomacro\gplfronttext{      \csname LTb\endcsname      \put(128,1292){\rotatebox{-270}{\makebox(0,0){\strut{}$\mathrm{v_r}$ ($c$)}}}      \put(2704,112){\makebox(0,0){\strut{}t ($\omega_{pi}^{-1}$)}}      \csname LTb\endcsname      \put(1936,975){\makebox(0,0)[r]{\strut{}$\Delta$t = 0.1}}      \csname LTb\endcsname      \put(1936,815){\makebox(0,0)[r]{\strut{}$\Delta$t = 0.001}}      \csname LTb\endcsname      \put(1936,655){\makebox(0,0)[r]{\strut{}Exact Sol.}}    }    \gplbacktext
    \put(0,0){\includegraphics{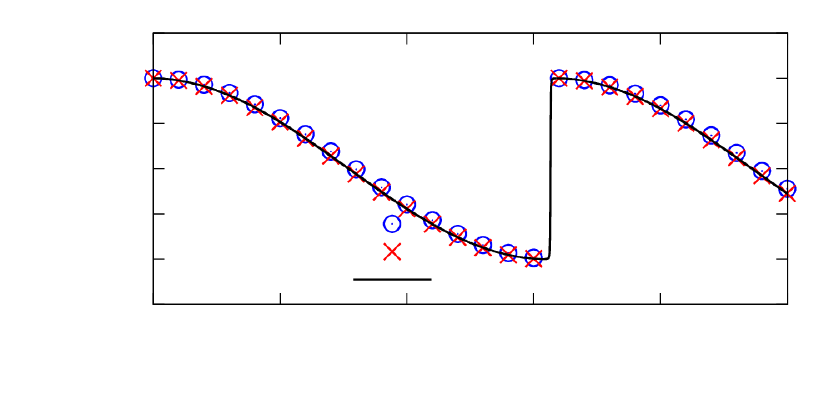}}    \gplfronttext
  \end{picture}\endgroup
   \caption[Radial component of the velocity]{Same as in Fig. \ref{figu.7}, but for the radial component of the velocity. The error for different $\Delta t$ is shown in \refFig{figu.9}} 
  \label{figu.8}
\end{figure}

The particle is advanced in time using Eqns. (\ref{eq_mover_v})--(\ref{eq_mover_r}). Then the result is compared with the solutions 
obtained by integrating the equations of motion in cylindrical geometry with the Matlab function {\tt ODE45}. The tolerance for the 
{\tt ODE45} solver is set to machine precision, so we can consider its result as the exact solution. Figures \ref{figu.7} 
and \ref{figu.8} show an example of the comparison between the ECsim-CYL mover and the Matlab solutions. 
In particular the evolution in time of the radial coordinate $r$ and the radial velocity $v_r$ are shown for two different time steps. 
In both cases the results from the code are in excellent agreement with the reference solution. 

\begin{figure}
  \center
  \begingroup
  \makeatletter
  \providecommand\color[2][]{    \GenericError{(gnuplot) \space\space\space\@spaces}{      Package color not loaded in conjunction with
      terminal option `colourtext'    }{See the gnuplot documentation for explanation.    }{Either use 'blacktext' in gnuplot or load the package
      color.sty in LaTeX.}    \renewcommand\color[2][]{}  }  \providecommand\includegraphics[2][]{    \GenericError{(gnuplot) \space\space\space\@spaces}{      Package graphicx or graphics not loaded    }{See the gnuplot documentation for explanation.    }{The gnuplot epslatex terminal needs graphicx.sty or graphics.sty.}    \renewcommand\includegraphics[2][]{}  }  \providecommand\rotatebox[2]{#2}  \@ifundefined{ifGPcolor}{    \newif\ifGPcolor
    \GPcolortrue
  }{}  \@ifundefined{ifGPblacktext}{    \newif\ifGPblacktext
    \GPblacktextfalse
  }{}  \let\gplgaddtomacro\g@addto@macro
  \gdef\gplbacktext{}  \gdef\gplfronttext{}  \makeatother
  \ifGPblacktext
    \def\colorrgb#1{}    \def\colorgray#1{}  \else
    \ifGPcolor
      \def\colorrgb#1{\color[rgb]{#1}}      \def\colorgray#1{\color[gray]{#1}}      \expandafter\def\csname LTw\endcsname{\color{white}}      \expandafter\def\csname LTb\endcsname{\color{black}}      \expandafter\def\csname LTa\endcsname{\color{black}}      \expandafter\def\csname LT0\endcsname{\color[rgb]{1,0,0}}      \expandafter\def\csname LT1\endcsname{\color[rgb]{0,1,0}}      \expandafter\def\csname LT2\endcsname{\color[rgb]{0,0,1}}      \expandafter\def\csname LT3\endcsname{\color[rgb]{1,0,1}}      \expandafter\def\csname LT4\endcsname{\color[rgb]{0,1,1}}      \expandafter\def\csname LT5\endcsname{\color[rgb]{1,1,0}}      \expandafter\def\csname LT6\endcsname{\color[rgb]{0,0,0}}      \expandafter\def\csname LT7\endcsname{\color[rgb]{1,0.3,0}}      \expandafter\def\csname LT8\endcsname{\color[rgb]{0.5,0.5,0.5}}    \else
      \def\colorrgb#1{\color{black}}      \def\colorgray#1{\color[gray]{#1}}      \expandafter\def\csname LTw\endcsname{\color{white}}      \expandafter\def\csname LTb\endcsname{\color{black}}      \expandafter\def\csname LTa\endcsname{\color{black}}      \expandafter\def\csname LT0\endcsname{\color{black}}      \expandafter\def\csname LT1\endcsname{\color{black}}      \expandafter\def\csname LT2\endcsname{\color{black}}      \expandafter\def\csname LT3\endcsname{\color{black}}      \expandafter\def\csname LT4\endcsname{\color{black}}      \expandafter\def\csname LT5\endcsname{\color{black}}      \expandafter\def\csname LT6\endcsname{\color{black}}      \expandafter\def\csname LT7\endcsname{\color{black}}      \expandafter\def\csname LT8\endcsname{\color{black}}    \fi
  \fi
    \setlength{\unitlength}{0.0500bp}    \ifx\gptboxheight\undefined      \newlength{\gptboxheight}      \newlength{\gptboxwidth}      \newsavebox{\gptboxtext}    \fi    \setlength{\fboxrule}{0.5pt}    \setlength{\fboxsep}{1pt}\begin{picture}(4818.00,2266.00)    \gplgaddtomacro\gplbacktext{      \csname LTb\endcsname      \put(784,512){\makebox(0,0)[r]{\strut{}1e-10}}      \put(784,735){\makebox(0,0)[r]{\strut{}1e-09}}      \put(784,958){\makebox(0,0)[r]{\strut{}1e-08}}      \put(784,1181){\makebox(0,0)[r]{\strut{}1e-07}}      \put(784,1404){\makebox(0,0)[r]{\strut{}1e-06}}      \put(784,1627){\makebox(0,0)[r]{\strut{}1e-05}}      \put(784,1850){\makebox(0,0)[r]{\strut{}1e-04}}      \put(784,2073){\makebox(0,0)[r]{\strut{}1e-03}}      \put(880,352){\makebox(0,0){\strut{}$0.001$}}      \put(3685,352){\makebox(0,0){\strut{}$0.01$}}    }    \gplgaddtomacro\gplfronttext{      \csname LTb\endcsname      \put(128,1292){\rotatebox{-270}{\makebox(0,0){\strut{}$\epsilon$}}}      \put(2704,112){\makebox(0,0){\strut{}$\Delta t$ ($\omega_{pi}^{-1}$)}}      \csname LTb\endcsname      \put(1264,1930){\makebox(0,0)[r]{\strut{}r}}      \csname LTb\endcsname      \put(1264,1770){\makebox(0,0)[r]{\strut{}$\mathrm{v}_r$}}    }    \gplbacktext
    \put(0,0){\includegraphics{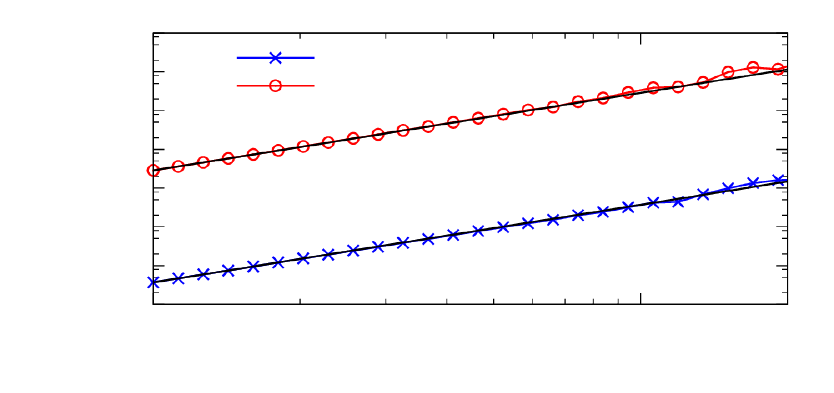}}    \gplfronttext
  \end{picture}\endgroup
   \caption[Error]{Error when evaluating the radial position (blue)  and velocity (red) for different $\Delta t$.
          Initial conditions of the particles and fields correspond to those in \refFig{figu.7}.
          The black lines represent the function $y=x^2$ to show the slope of the error.} 
  \label{figu.9}
\end{figure}

We have also carried out tests where we vary the time step used to compute the particle trajectories with the ECsim-CYL mover, in order to study the influence on overall accuracy of this parameter.
The error on the radial position $\epsilon_r$ was computed as:
\begin{equation}
\epsilon_r = \frac{\delta_r}{M_r},
\end{equation}
where $\delta_r$ is the variance. $\delta_r = \int ( r_{ODE} - r_{ECSIM} )^2 dt$, and $M_r$ is the average value, $M_r = \int r^2_{ODE} dt$. 
The same definitions hold for the error on the particle radial velocity. Results are shown in Fig. \ref{figu.9}, in which $\epsilon_r$ and $\epsilon_{v_r}$ are plotted for different values of $\Delta t$. The error on the particle orbit increases quadratically with $\Delta t$, thus confirming that the mover is second order accurate in time. 

\subsection{Field solver}
\begin{figure}
  \center
  \begingroup
  \makeatletter
  \providecommand\color[2][]{    \GenericError{(gnuplot) \space\space\space\@spaces}{      Package color not loaded in conjunction with
      terminal option `colourtext'    }{See the gnuplot documentation for explanation.    }{Either use 'blacktext' in gnuplot or load the package
      color.sty in LaTeX.}    \renewcommand\color[2][]{}  }  \providecommand\includegraphics[2][]{    \GenericError{(gnuplot) \space\space\space\@spaces}{      Package graphicx or graphics not loaded    }{See the gnuplot documentation for explanation.    }{The gnuplot epslatex terminal needs graphicx.sty or graphics.sty.}    \renewcommand\includegraphics[2][]{}  }  \providecommand\rotatebox[2]{#2}  \@ifundefined{ifGPcolor}{    \newif\ifGPcolor
    \GPcolortrue
  }{}  \@ifundefined{ifGPblacktext}{    \newif\ifGPblacktext
    \GPblacktextfalse
  }{}  \let\gplgaddtomacro\g@addto@macro
  \gdef\gplbacktext{}  \gdef\gplfronttext{}  \makeatother
  \ifGPblacktext
    \def\colorrgb#1{}    \def\colorgray#1{}  \else
    \ifGPcolor
      \def\colorrgb#1{\color[rgb]{#1}}      \def\colorgray#1{\color[gray]{#1}}      \expandafter\def\csname LTw\endcsname{\color{white}}      \expandafter\def\csname LTb\endcsname{\color{black}}      \expandafter\def\csname LTa\endcsname{\color{black}}      \expandafter\def\csname LT0\endcsname{\color[rgb]{1,0,0}}      \expandafter\def\csname LT1\endcsname{\color[rgb]{0,1,0}}      \expandafter\def\csname LT2\endcsname{\color[rgb]{0,0,1}}      \expandafter\def\csname LT3\endcsname{\color[rgb]{1,0,1}}      \expandafter\def\csname LT4\endcsname{\color[rgb]{0,1,1}}      \expandafter\def\csname LT5\endcsname{\color[rgb]{1,1,0}}      \expandafter\def\csname LT6\endcsname{\color[rgb]{0,0,0}}      \expandafter\def\csname LT7\endcsname{\color[rgb]{1,0.3,0}}      \expandafter\def\csname LT8\endcsname{\color[rgb]{0.5,0.5,0.5}}    \else
      \def\colorrgb#1{\color{black}}      \def\colorgray#1{\color[gray]{#1}}      \expandafter\def\csname LTw\endcsname{\color{white}}      \expandafter\def\csname LTb\endcsname{\color{black}}      \expandafter\def\csname LTa\endcsname{\color{black}}      \expandafter\def\csname LT0\endcsname{\color{black}}      \expandafter\def\csname LT1\endcsname{\color{black}}      \expandafter\def\csname LT2\endcsname{\color{black}}      \expandafter\def\csname LT3\endcsname{\color{black}}      \expandafter\def\csname LT4\endcsname{\color{black}}      \expandafter\def\csname LT5\endcsname{\color{black}}      \expandafter\def\csname LT6\endcsname{\color{black}}      \expandafter\def\csname LT7\endcsname{\color{black}}      \expandafter\def\csname LT8\endcsname{\color{black}}    \fi
  \fi
    \setlength{\unitlength}{0.0500bp}    \ifx\gptboxheight\undefined      \newlength{\gptboxheight}      \newlength{\gptboxwidth}      \newsavebox{\gptboxtext}    \fi    \setlength{\fboxrule}{0.5pt}    \setlength{\fboxsep}{1pt}\begin{picture}(4818.00,2834.00)    \gplgaddtomacro\gplbacktext{      \csname LTb\endcsname      \put(688,645){\makebox(0,0)[r]{\strut{}$-0.6$}}      \put(688,911){\makebox(0,0)[r]{\strut{}$-0.4$}}      \put(688,1177){\makebox(0,0)[r]{\strut{}$-0.2$}}      \put(688,1443){\makebox(0,0)[r]{\strut{}$0$}}      \put(688,1710){\makebox(0,0)[r]{\strut{}$0.2$}}      \put(688,1976){\makebox(0,0)[r]{\strut{}$0.4$}}      \put(688,2242){\makebox(0,0)[r]{\strut{}$0.6$}}      \put(688,2508){\makebox(0,0)[r]{\strut{}$0.8$}}      \put(784,352){\makebox(0,0){\strut{}$0$}}      \put(1380,352){\makebox(0,0){\strut{}$1$}}      \put(1976,352){\makebox(0,0){\strut{}$2$}}      \put(2572,352){\makebox(0,0){\strut{}$3$}}      \put(3168,352){\makebox(0,0){\strut{}$4$}}      \put(3764,352){\makebox(0,0){\strut{}$5$}}      \put(4360,352){\makebox(0,0){\strut{}$6$}}    }    \gplgaddtomacro\gplfronttext{      \csname LTb\endcsname      \put(128,1576){\rotatebox{-270}{\makebox(0,0){\strut{}$E_\phi$}}}      \put(2656,112){\makebox(0,0){\strut{}r ($d_i$)}}      \csname LTb\endcsname      \put(3794,2498){\makebox(0,0)[r]{\strut{}Analytical}}      \csname LTb\endcsname      \put(3794,2338){\makebox(0,0)[r]{\strut{}ECsimCYL}}    }    \gplbacktext
    \put(0,0){\includegraphics{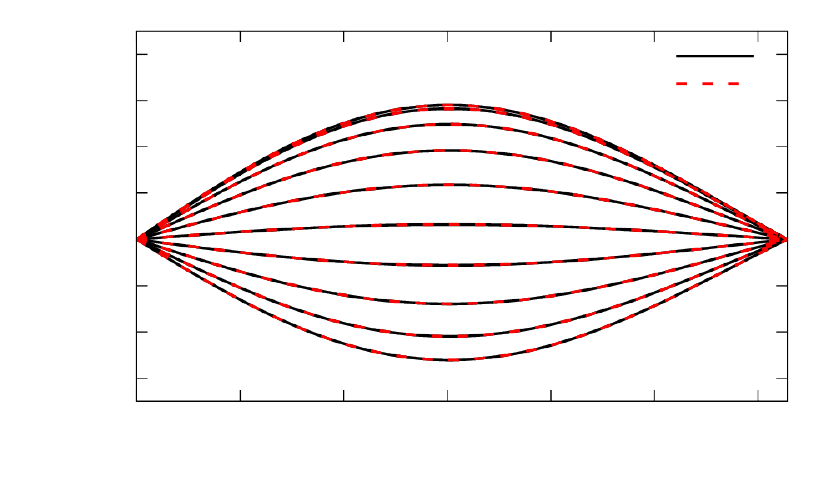}}    \gplfronttext
  \end{picture}\endgroup
   \caption[Comparison between numerical and analytical $E_{\phi}$]{Comparison between numerical and analytical $E_{\phi}$ 
           every $0.5 \, \omega_{pi}^{-1}$ for the last $5 \, \omega_{pi}^{-1}$. Black solid
  lines represent the reference solutions, while red dashed lines the numerical ones.}\label{fig.9}
\end{figure}

The cylindrical electromagnetic wave propagation in vacuum is studied with a 
one-dimensional version of the field solver presented in section \ref{sec.2.1}. We investigate the TE (Transverse Electric, or $E_z = 0$) mode. 
The problem presents an analytical solution, given by: 

\[
   \begin{dcases}
     B_z = -J_0(\omega_n r) \sin(\omega_n t), \\ 
     E_{\phi} = J_1 (\omega_n r) \cos(\omega_n t),
   \end{dcases}
\]
where $J_0$ denotes the Bessel function of the first kind of order 0, $J_1$ the Bessel function
of the first kind of order 1 and $\omega_n = z_1/L$, with $z_1$ the zero of the Bessel function of the first kind and
order 1 and $L$ the length of the domain.
Therefore in order to analyze the propagation of 
the electromagnetic fields, the following initial conditions for the electric 
and magnetic fields are considered in the code: $\mathbf{E} = E_{\phi} 
\hat{\boldsymbol{\phi}}$, with $E_{\phi} = J_1 (\omega_n r)$ and $\mathbf{B} = 0$. 
A comparison between analytical and numerical solutions is reported in Fig. \ref{fig.9}, which 
show the fields as a function of $r$ at different times. 
The plots are obtained choosing the size of the domain equal to $L = 2 \pi$, using 128 cells and a 
time step equal to $\Delta t = 0.01$. The duration of the runs is $t_{i\text{end}}=8\pi \omega_{pi}^{-1}$. 
The curves are in excellent agreement.

\begin{figure}
  \center
  \begingroup
  \makeatletter
  \providecommand\color[2][]{    \GenericError{(gnuplot) \space\space\space\@spaces}{      Package color not loaded in conjunction with
      terminal option `colourtext'    }{See the gnuplot documentation for explanation.    }{Either use 'blacktext' in gnuplot or load the package
      color.sty in LaTeX.}    \renewcommand\color[2][]{}  }  \providecommand\includegraphics[2][]{    \GenericError{(gnuplot) \space\space\space\@spaces}{      Package graphicx or graphics not loaded    }{See the gnuplot documentation for explanation.    }{The gnuplot epslatex terminal needs graphicx.sty or graphics.sty.}    \renewcommand\includegraphics[2][]{}  }  \providecommand\rotatebox[2]{#2}  \@ifundefined{ifGPcolor}{    \newif\ifGPcolor
    \GPcolortrue
  }{}  \@ifundefined{ifGPblacktext}{    \newif\ifGPblacktext
    \GPblacktextfalse
  }{}  \let\gplgaddtomacro\g@addto@macro
  \gdef\gplbacktext{}  \gdef\gplfronttext{}  \makeatother
  \ifGPblacktext
    \def\colorrgb#1{}    \def\colorgray#1{}  \else
    \ifGPcolor
      \def\colorrgb#1{\color[rgb]{#1}}      \def\colorgray#1{\color[gray]{#1}}      \expandafter\def\csname LTw\endcsname{\color{white}}      \expandafter\def\csname LTb\endcsname{\color{black}}      \expandafter\def\csname LTa\endcsname{\color{black}}      \expandafter\def\csname LT0\endcsname{\color[rgb]{1,0,0}}      \expandafter\def\csname LT1\endcsname{\color[rgb]{0,1,0}}      \expandafter\def\csname LT2\endcsname{\color[rgb]{0,0,1}}      \expandafter\def\csname LT3\endcsname{\color[rgb]{1,0,1}}      \expandafter\def\csname LT4\endcsname{\color[rgb]{0,1,1}}      \expandafter\def\csname LT5\endcsname{\color[rgb]{1,1,0}}      \expandafter\def\csname LT6\endcsname{\color[rgb]{0,0,0}}      \expandafter\def\csname LT7\endcsname{\color[rgb]{1,0.3,0}}      \expandafter\def\csname LT8\endcsname{\color[rgb]{0.5,0.5,0.5}}    \else
      \def\colorrgb#1{\color{black}}      \def\colorgray#1{\color[gray]{#1}}      \expandafter\def\csname LTw\endcsname{\color{white}}      \expandafter\def\csname LTb\endcsname{\color{black}}      \expandafter\def\csname LTa\endcsname{\color{black}}      \expandafter\def\csname LT0\endcsname{\color{black}}      \expandafter\def\csname LT1\endcsname{\color{black}}      \expandafter\def\csname LT2\endcsname{\color{black}}      \expandafter\def\csname LT3\endcsname{\color{black}}      \expandafter\def\csname LT4\endcsname{\color{black}}      \expandafter\def\csname LT5\endcsname{\color{black}}      \expandafter\def\csname LT6\endcsname{\color{black}}      \expandafter\def\csname LT7\endcsname{\color{black}}      \expandafter\def\csname LT8\endcsname{\color{black}}    \fi
  \fi
    \setlength{\unitlength}{0.0500bp}    \ifx\gptboxheight\undefined      \newlength{\gptboxheight}      \newlength{\gptboxwidth}      \newsavebox{\gptboxtext}    \fi    \setlength{\fboxrule}{0.5pt}    \setlength{\fboxsep}{1pt}\begin{picture}(4818.00,2834.00)    \gplgaddtomacro\gplbacktext{      \csname LTb\endcsname      \put(688,512){\makebox(0,0)[r]{\strut{}$-1.2$}}      \put(688,762){\makebox(0,0)[r]{\strut{}$-1$}}      \put(688,1013){\makebox(0,0)[r]{\strut{}$-0.8$}}      \put(688,1263){\makebox(0,0)[r]{\strut{}$-0.6$}}      \put(688,1514){\makebox(0,0)[r]{\strut{}$-0.4$}}      \put(688,1764){\makebox(0,0)[r]{\strut{}$-0.2$}}      \put(688,2015){\makebox(0,0)[r]{\strut{}$0$}}      \put(688,2265){\makebox(0,0)[r]{\strut{}$0.2$}}      \put(688,2516){\makebox(0,0)[r]{\strut{}$0.4$}}      \put(784,352){\makebox(0,0){\strut{}$0$}}      \put(1380,352){\makebox(0,0){\strut{}$1$}}      \put(1976,352){\makebox(0,0){\strut{}$2$}}      \put(2572,352){\makebox(0,0){\strut{}$3$}}      \put(3168,352){\makebox(0,0){\strut{}$4$}}      \put(3764,352){\makebox(0,0){\strut{}$5$}}      \put(4360,352){\makebox(0,0){\strut{}$6$}}    }    \gplgaddtomacro\gplfronttext{      \csname LTb\endcsname      \put(128,1576){\rotatebox{-270}{\makebox(0,0){\strut{}$B_z$}}}      \put(2656,112){\makebox(0,0){\strut{}r ($d_i$)}}      \csname LTb\endcsname      \put(1840,2498){\makebox(0,0)[r]{\strut{}Analytical}}      \csname LTb\endcsname      \put(1840,2338){\makebox(0,0)[r]{\strut{}ECsimCYL}}    }    \gplbacktext
    \put(0,0){\includegraphics{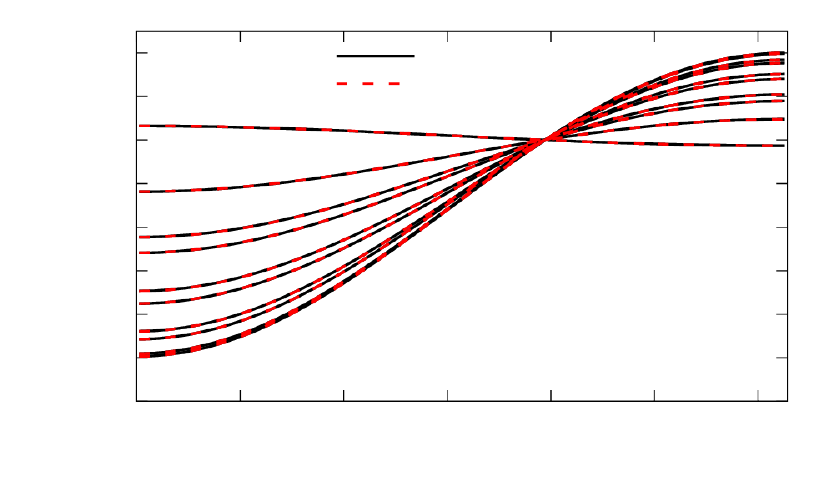}}    \gplfronttext
  \end{picture}\endgroup
   \caption[Comparison between numerical and analytical $B_z$]{Same as in Fig. \ref{fig.9}, 
  but for the z-component of the magnetic field.}\label{fig.10}
\end{figure}

\begin{figure}
  \center
  \begingroup
  \makeatletter
  \providecommand\color[2][]{    \GenericError{(gnuplot) \space\space\space\@spaces}{      Package color not loaded in conjunction with
      terminal option `colourtext'    }{See the gnuplot documentation for explanation.    }{Either use 'blacktext' in gnuplot or load the package
      color.sty in LaTeX.}    \renewcommand\color[2][]{}  }  \providecommand\includegraphics[2][]{    \GenericError{(gnuplot) \space\space\space\@spaces}{      Package graphicx or graphics not loaded    }{See the gnuplot documentation for explanation.    }{The gnuplot epslatex terminal needs graphicx.sty or graphics.sty.}    \renewcommand\includegraphics[2][]{}  }  \providecommand\rotatebox[2]{#2}  \@ifundefined{ifGPcolor}{    \newif\ifGPcolor
    \GPcolortrue
  }{}  \@ifundefined{ifGPblacktext}{    \newif\ifGPblacktext
    \GPblacktextfalse
  }{}  \let\gplgaddtomacro\g@addto@macro
  \gdef\gplbacktext{}  \gdef\gplfronttext{}  \makeatother
  \ifGPblacktext
    \def\colorrgb#1{}    \def\colorgray#1{}  \else
    \ifGPcolor
      \def\colorrgb#1{\color[rgb]{#1}}      \def\colorgray#1{\color[gray]{#1}}      \expandafter\def\csname LTw\endcsname{\color{white}}      \expandafter\def\csname LTb\endcsname{\color{black}}      \expandafter\def\csname LTa\endcsname{\color{black}}      \expandafter\def\csname LT0\endcsname{\color[rgb]{1,0,0}}      \expandafter\def\csname LT1\endcsname{\color[rgb]{0,1,0}}      \expandafter\def\csname LT2\endcsname{\color[rgb]{0,0,1}}      \expandafter\def\csname LT3\endcsname{\color[rgb]{1,0,1}}      \expandafter\def\csname LT4\endcsname{\color[rgb]{0,1,1}}      \expandafter\def\csname LT5\endcsname{\color[rgb]{1,1,0}}      \expandafter\def\csname LT6\endcsname{\color[rgb]{0,0,0}}      \expandafter\def\csname LT7\endcsname{\color[rgb]{1,0.3,0}}      \expandafter\def\csname LT8\endcsname{\color[rgb]{0.5,0.5,0.5}}    \else
      \def\colorrgb#1{\color{black}}      \def\colorgray#1{\color[gray]{#1}}      \expandafter\def\csname LTw\endcsname{\color{white}}      \expandafter\def\csname LTb\endcsname{\color{black}}      \expandafter\def\csname LTa\endcsname{\color{black}}      \expandafter\def\csname LT0\endcsname{\color{black}}      \expandafter\def\csname LT1\endcsname{\color{black}}      \expandafter\def\csname LT2\endcsname{\color{black}}      \expandafter\def\csname LT3\endcsname{\color{black}}      \expandafter\def\csname LT4\endcsname{\color{black}}      \expandafter\def\csname LT5\endcsname{\color{black}}      \expandafter\def\csname LT6\endcsname{\color{black}}      \expandafter\def\csname LT7\endcsname{\color{black}}      \expandafter\def\csname LT8\endcsname{\color{black}}    \fi
  \fi
    \setlength{\unitlength}{0.0500bp}    \ifx\gptboxheight\undefined      \newlength{\gptboxheight}      \newlength{\gptboxwidth}      \newsavebox{\gptboxtext}    \fi    \setlength{\fboxrule}{0.5pt}    \setlength{\fboxsep}{1pt}\begin{picture}(4818.00,2266.00)    \gplgaddtomacro\gplbacktext{      \csname LTb\endcsname      \put(784,512){\makebox(0,0)[r]{\strut{}1e-06}}      \put(784,824){\makebox(0,0)[r]{\strut{}1e-05}}      \put(784,1136){\makebox(0,0)[r]{\strut{}1e-04}}      \put(784,1449){\makebox(0,0)[r]{\strut{}1e-03}}      \put(784,1761){\makebox(0,0)[r]{\strut{}1e-02}}      \put(784,2073){\makebox(0,0)[r]{\strut{}1e-01}}      \put(880,352){\makebox(0,0){\strut{}$0.001$}}      \put(2704,352){\makebox(0,0){\strut{}$0.01$}}      \put(4529,352){\makebox(0,0){\strut{}$0.1$}}    }    \gplgaddtomacro\gplfronttext{      \csname LTb\endcsname      \put(128,1292){\rotatebox{-270}{\makebox(0,0){\strut{}$\epsilon$}}}      \put(2704,112){\makebox(0,0){\strut{}$\Delta t$ ($\omega_{pi}^{-1}$)}}      \csname LTb\endcsname      \put(1264,1930){\makebox(0,0)[r]{\strut{}$\mathrm{E_\phi}$}}      \csname LTb\endcsname      \put(1264,1770){\makebox(0,0)[r]{\strut{}$\mathrm{B_z}$}}    }    \gplbacktext
    \put(0,0){\includegraphics{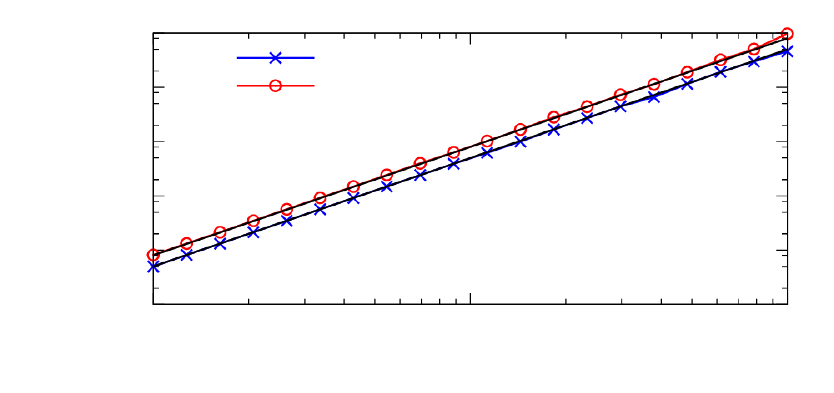}}    \gplfronttext
  \end{picture}\endgroup
   \caption[Evolution of the error on $E_{\phi}$ and $B_z$ with respect to $\Delta t$]{Error in 
  evaluating $E_\phi$ and $B_z$ for different $\Delta t$ and $\text{CFL}=0.1$. The 
  black line represents the function $y=x^2$ to show the slope of the 
  error.}\label{figur.11}
\end{figure}

The numerical error introduced by the code has then been computed after the 
numerical solution was advanced until a chosen time step $t_{\text{end}}$. The error on 
a given field $F(r,t_{\text{end}})$ is computed as:
\begin{equation}
  \epsilon_{F} = \frac{||F^{\text{an}}(r,t_{\text{end}}) -F^{\text{num}}(r,t_{\text{end}}) 
  ||}{||F^{\text{an}}(r,t_{\text{end}})||},
\end{equation}
where the superscripts $\text{an}$ and $\text{num}$ refer to the analytical and numerical 
solutions at $t=t_{\text{end}}$. The quantities $|| x(r,t_{\text{end}}) ||$ indicate the 
Euclidean norm, computed as
\begin{equation}
  ||x|| = \sqrt{\sum_{r_i=1}^{N_r} x(r_i)^2} ,
\end{equation}
where the summation is over the $N_r$ radial points $r_i$ into which the domain 
is discretize. Results are presented in Fig.  \ref{figur.11}, which shows the 
error introduced by the code for different values 
of $\Delta t$, assuming $\text{CFL}=\Delta t / \Delta x$ constant (quantities in 
dimensionless units). Both in the case of $E_{\phi}$  and $B_z$, the error 
varies quadratically with $\Delta t$ indicating that also the field solver is a 
second order scheme in time.

\section{Plasma wave spectrum in homogeneous plasma}

This test case consists of an homogeneous electron-positron plasma which completely
fills a cylindrical waveguide. Electrons and positrons have a Maxwellian distribution
for the velocities and they are at the same temperature. 
This situation is well known and according to electromagnetic theory, in a cylindrical 
waveguide filled with an homogeneous plasma there is one solution of the Maxwell's equations called the "magnetic 
mode" or "H wave" \citep{Khalil2014}. These waves have a dispersion relation given by
\begin{equation}
  k = \frac{\omega}{c}\sqrt{1 - \left(\frac{\omega_c}{\omega}\right)^2},
  \label{eq-disprel}
\end{equation}
where $\omega_c$ is the cutoff frequency.
\begin{equation}
  \omega_c = \sqrt{\omega_{pi} + \left(\frac{c \chi_{\ell,m}}{R}\right)}
  \label{eq-cutoff}
\end{equation}
In \refEqu{eq-cutoff} $\omega_p$ is the plasma frequency, $\chi_{\ell,m}$ the root $m$ of the Bessel functions $J_{\ell}$, 
$c$ the speed of light and $R$ the radius of the guide. In our case, since the system is axially symmetric,
only the wave with $\ell = 0$ exists. In order to validate the code, we have designed a case which reproduces
this situation and we have analyzed the $k-\omega$ spectra.

The size of the domain is determined by its radius $R=10 d_i$ and its length $L=20 d_i$. We have $100$ cells in $r$ 
direction and $200$ in $z$ direction.  We have used 1 000 000 particles with a thermal velocity of 
$v_{\text{th}} = 0.01 \, c$, 500 000 electrons and 500 000 positrons. Both sets of particles are initialized randomly 
in the entire domain. The time step is $\Delta t = 0.005 \omega_p^{-1}$ and we have run the simulation for
5 000 cycles (25 $\omega_p^{-1}$). We have assumed that the waveguide is a perfect conductor, hence at the
outer boundary conditions both the tangential electric components $E_z$ and $E_\phi$ and the derivative of the
 normal electric component $E_r$ are nil. 

\begin{figure}[h]
  \center
  \includegraphics[width=3.0in]{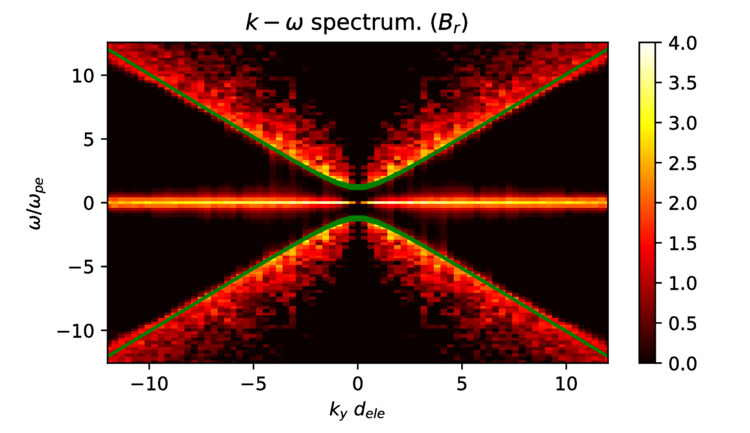}
  \caption{$k-\omega$ spectrum of the radial component of the magnetic field (in
           logarithmic scale). In green the theoretical dispersion relation ---\refEqu{eq-disprel}--- of the H waves.}
  \label{fig-spectra}
\end{figure}

In \refFig{fig-spectra} we show the $k-\omega$ spectrum of the simulation
together with the dispersion relation given by \refEqu{eq-disprel} for $m = 1,2,3,4,5$ (green line). As the abscissa we have
the $k$ number in the $z$ direction in units of ion skin depth ($k_z d_i$) and as the ordinate the frequency
in units of inverse of the plasma frequency ($\omega/\omega_{pi}$). The simulation results and the theory are in
very good agreement. As should be expected, at large $k$ and $\omega$, the numerical discretization of the operators introduces a (second-order, as shown above) truncation error and the results from the simulations start to diverge slightly 
from the analytical solution. 

\section{Free Expansion}

In order to further test the accuracy and capability of the ECsim-CYL implementation, we have
investigated homogeneous expansion of a plasma sphere in vacuum.  The test covers the following three
areas: 1) How ECsim-CYL preserves spherical symmetry with the use of cylindrical coordinates, 2) How
ECsim-CYL conserves total energy of the system, and 3) How ECsim-CYL describes ambipolar plasma
diffusion with respect to charge separation, self-consistent ambipolar electric field, and change in
velocity distribution of ions and electrons.  The underlying mechanism of ambipolar plasma diffusion
lies with the difference in the particle velocity between lighter electrons and heavier ions.
Initially, electrons will expand faster than ions due to their lighter mass and faster velocity for
the same temperature.  This leads to charge separation between electrons and ions, which in turn
generates an electric field.  The spatio-temporal profile and the magnitude of this electric field is then
adjusted by the plasma motion as the electric field slows down the rate of electron expansion and
accelerates the rate of ion expansion until the rate of ion and electron expansion is equalized.
This phenomenon, called ambipolar diffusion, is well known in plasma physics and further details can
be found in many standard textbooks \cite{KrallTrivelpiece}.

\begin{table}[h]
  \center {
  \begin{tabular}{|l|cccccccc|}
  \hline
   Case  & $n_i/m_e$        &  \# cells       & $\Delta x$/$d_i$ & $\Delta x/\lambda_D$  & $\Delta t \, \omega_{pi}$ & Cycles & \# part          & CPU hours   \\
  \hline                                                                                                                  
   A     &  25     & $25\times50  $           &  0.2             &  10.0                & 0.4                       & 100     & $6.6\cdot10^4$ & 0.08  \\
   B     &  25     & $50\times100 $           &  0.1             &  5.0                 & 0.2                       & 200     & $2.7\cdot10^5$ & 0.65  \\
   C     &  25     & $100\times200$           &  0.05            &  2.5                 & 0.1                       & 400     & $1.0\cdot10^6$ & 5.47  \\
   D     &  25     & $250\times500$           &  0.02            &  1.0                 & 0.04                      & 1000    & $6.6\cdot10^6$ & 88.33 \\
   A-3D  &  25     & $50 \times 50 \times50$  &  0.2             &  10.0                & 0.4                       & 100     & $6.6\cdot10^6$ & 3.97  \\
   A-1836  &  1836   & $25\times50  $           &  0.2             &  10.0                & 0.4                       & 857     & $6.6\cdot10^4$ & 0.70  \\
   D-1836  &  1836   & $250\times500$           &  0.02            &  1.0                 & 0.04                      & 8570    & $6.6\cdot10^6$ & 759.81 \\
  \hline
  \end{tabular}
  \caption{Simulation parameters used for the test cases in Section 5.}
  \label{tab-free}
  }
\end{table}

\refTab{tab-free} summarizes the simulation conditions used in the test.  For a majority of the tests, 
we have used light ions with $m_i  = 25\; m_e$ with the same temperature for electrons and ions.  This reduced 
ion mass is used to speed up the computation time compared to the realistic ion mass of $1836\; m_e$.  Later, 
we will present the result for the hydrogen ion with the mass of $1836\; m_e$ and compare the results with
the reduced mass case of $m_i  = 25\; m_e$.  The simulation begins with an initial homogeneous plasma
inside a sphere with a radius of $R_s$ with Maxwellian distribution for ions and electrons with an
electron thermal velocity of $v_{\text{th},\text{e}}^r = v_{\text{th},\text{e}}^\theta = v_{\text{th},\text{e}}^z = 0.1 \, c$ and and an ion thermal velocity of
$v_{\text{th},\text{i}}^r = v_{\text{th},\text{i}}^\theta = v_{\text{th},\text{i}}^z = 0.02 \, c$
where $c$ is the speed of light.  It is noted that the ECsim-CYL like ECsim and iPic3D utilizes the normalized unit with c = 1.  

The initial charge particle density is set inside a sphere for $R_s =
1\; d_i$ or $R_s = 50 \; \lambda_D$, where $d_i = c/\omega_{pi}$ is the ion inertial length,
$\omega_{pi}$ is the ion plasma frequency, $\lambda_D = v_{\text{th},\text{e}}/\omega_{pe}$
is the electron Debye length, and $\omega_{pe}$ is the electron plasma frequency.  

The radius and
length of the cylindrical simulation domain are $R = 5\; d_i$ and $L  = 10\;  d_i$. As for the time step, the
Courant-Friedrichs-Lewy (CFL) condition of $v_{\text{th},\text{e}} \cdot \Delta t / \Delta x = 0.2$ 
is utilized for all cases in \refTab{tab-free}. As a semi-implicit particle code, this CFL condition is the only 
numerical constraint of ECsim-CYL regarding the size of time step\citep{lapenta2012particle,lapenta2016exactly}.  In addition, the total number of particles in each case is
chosen to keep the particle density per cell constant during initialization, while all simulation particles have the same charge except for the sign. 
Since all the particles have the same weight, to achieve a uniform charge distribution the number of particles per cell depends on the volume of each cell, 
which depends on the radial distance to the axis.

The calculations have been done using 16 Intel Xeon E5-2643 3.30 GHz cores. In some of the smaller cases only 2 or 4 cores have
been used. The amount of CPU hours of each case is the results of the number of cores used multiplied by the total runtime.

\subsection{Spherical symmetry}
\begin{figure}[h]
  \center
  \includegraphics[width=3.0in]{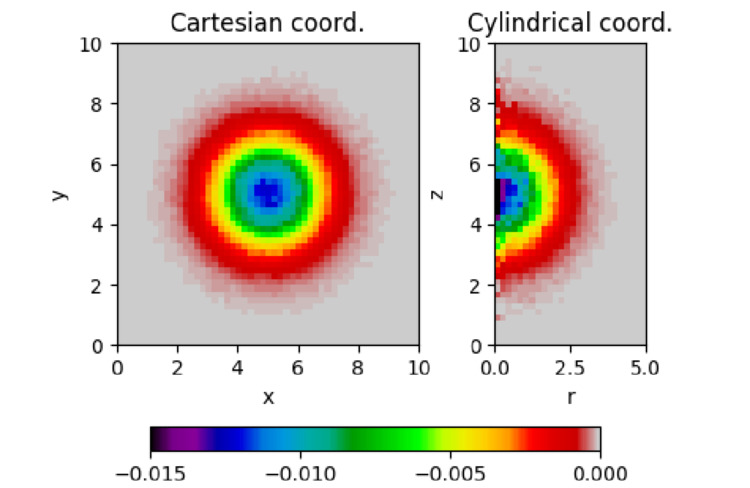}
  \caption{Comparison of electron charge density profile at $t=40 /\omega_{pi}$ Left from case A-3D using the 3D Cartesian ECsim code and Right - from case A using the cylindrical ECsim-CYL code.}
  \label{fig-comp3D2D}
\end{figure}

Since the code must reproduce the plasma dynamics independent of its coordinate choice, we conducted
homogeneous expansion of a plasma sphere in vacuum with the 3D Cartesian ECsim and with ECsim-CYL.  \refFig{fig-comp3D2D} compares the
electron density profiles on XY plane at $t=40 /\omega_{pi}$ from the case A-3D using ECsim and the case A using ECsim-CYL,
which shows good agreement between the two with well preserved spherical symmetry.  
As shown in \refTab{tab-free}, the case A-3D run using ECsim utilizes $50
\times 50 \times 50$ cells with $6.6 \cdot 10^6$ particles.  In comparison, the case A using 
ECsim-CYL utilizes $25 \times 50 \times 1$ cells with $6.6 \cdot 10^4$ particles.  These two runs use the 
same time step of $0.4/\omega_{pi}$ and the same spatial resolution in the radial and vertical
direction of $0.2 \; d_i$. In terms of computational cost, the case A run took only 0.08 CPU hours 
to complete 100 cycles compared to 4 CPU hours for the case A-3D, an improvement by a factor of 50. 
This improvement in speed is related to the reduction in dimensionality by taking an advantage of axisymmetric condition, 
validating our motivation to develop the ECsim-CYL. This gain in speed will be even more significant with increasing
simulation scale as we estimate that the ECsim-CYL would be about 500 times faster than the ECsim in
the case of 250 x 500 grid resolution.

\begin{figure}[h]
  \center
  \includegraphics[width=3.0in]{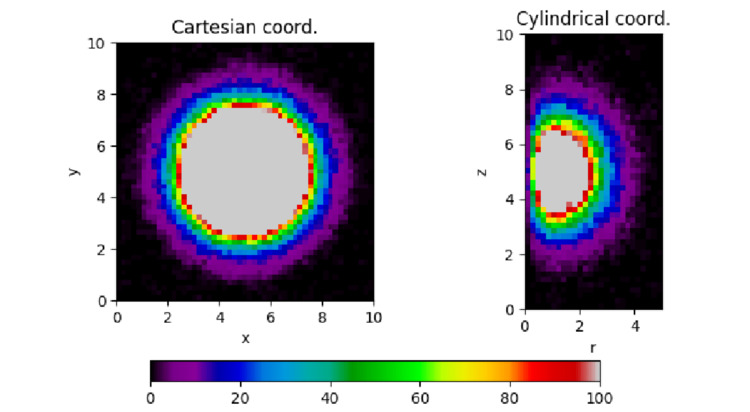}
  \caption{Comparison of number of particles per cell profile at $t=40 /\omega_{pi}$, Left Ð from case A-3D and right from case A.}
  \label{fig-comp3D2D_N}
\end{figure}

\begin{figure}[h]
  \center
  \includegraphics[width=3.0in]{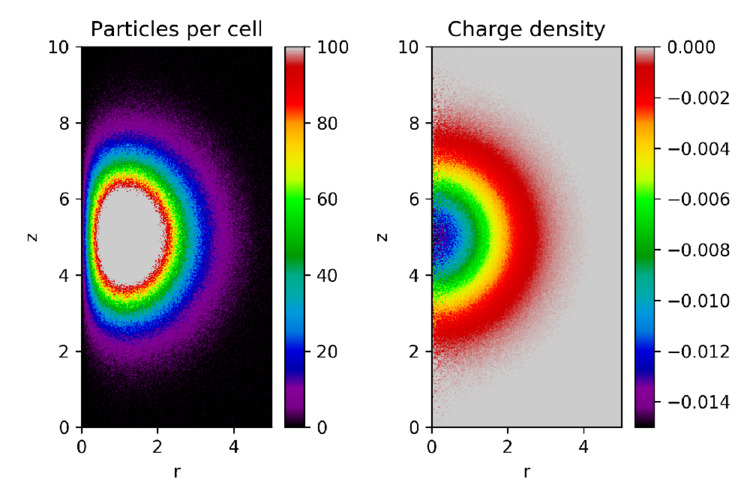}
  \caption{Number of particles per cell (left)  and electron charge density (right) at $t=40 /\omega_{pi}$ from case D}
  \label{fig-caseD_N_rho}
\end{figure}

It is also noted that there are some differences between the results of ECsim-CYL and ECsim as
understanding these differences would be important in utilizing ECsim-CYL.  The electron density profile
from ECsim-CYL in \refFig{fig-comp3D2D}  exhibits significant noise near the axis, especially for the 1st cells in the radial direction.  
This is expected for a cylindrical code as the volume of the cell
becomes very small near r=0 and contains significantly fewer particles per cell as shown in
\refFig{fig-comp3D2D_N}.  On the other and, the particle number per cell increases with increasing radius in ECsim-CYL.  Outside
r=1, the particle number density becomes comparable between the ECsim run and the ECsim-CYL run leading to a good agreement between them 
despite the fact that ECsim-CYL utilizes 100 times less particles.  As such, the ECsim-CYL is an excellent tool to
investigate the plasma dynamics of the axisymmetric system where the main region of interest is
away from the axis of symmetry.  On the other hand, caution must be taken to ensure the accuracy
of results near the axis for a system where the accurate description of plasma motion near the axis is important.  
One way to mitigate the axis problem is to increase the number of cells in the simulation.  As shown in \refFig{fig-caseD_N_rho} from the case D run,
the use of a large number of cells greatly reduces
the noise near the axis and provides improved spatial profiles albeit at a computational cost.

\subsection{Energy Conservation}

The next test is the energy conservation of ECsim-CYL.  Previously, we developed the ECsim code to
improve the stability of semi-implicit code at large spatial scales by implementing energy
conserving algorithm.

\begin{figure}[h]
  \center
  \begingroup
  \makeatletter
  \providecommand\color[2][]{    \GenericError{(gnuplot) \space\space\space\@spaces}{      Package color not loaded in conjunction with
      terminal option `colourtext'    }{See the gnuplot documentation for explanation.    }{Either use 'blacktext' in gnuplot or load the package
      color.sty in LaTeX.}    \renewcommand\color[2][]{}  }  \providecommand\includegraphics[2][]{    \GenericError{(gnuplot) \space\space\space\@spaces}{      Package graphicx or graphics not loaded    }{See the gnuplot documentation for explanation.    }{The gnuplot epslatex terminal needs graphicx.sty or graphics.sty.}    \renewcommand\includegraphics[2][]{}  }  \providecommand\rotatebox[2]{#2}  \@ifundefined{ifGPcolor}{    \newif\ifGPcolor
    \GPcolortrue
  }{}  \@ifundefined{ifGPblacktext}{    \newif\ifGPblacktext
    \GPblacktextfalse
  }{}  \let\gplgaddtomacro\g@addto@macro
  \gdef\gplbacktext{}  \gdef\gplfronttext{}  \makeatother
  \ifGPblacktext
    \def\colorrgb#1{}    \def\colorgray#1{}  \else
    \ifGPcolor
      \def\colorrgb#1{\color[rgb]{#1}}      \def\colorgray#1{\color[gray]{#1}}      \expandafter\def\csname LTw\endcsname{\color{white}}      \expandafter\def\csname LTb\endcsname{\color{black}}      \expandafter\def\csname LTa\endcsname{\color{black}}      \expandafter\def\csname LT0\endcsname{\color[rgb]{1,0,0}}      \expandafter\def\csname LT1\endcsname{\color[rgb]{0,1,0}}      \expandafter\def\csname LT2\endcsname{\color[rgb]{0,0,1}}      \expandafter\def\csname LT3\endcsname{\color[rgb]{1,0,1}}      \expandafter\def\csname LT4\endcsname{\color[rgb]{0,1,1}}      \expandafter\def\csname LT5\endcsname{\color[rgb]{1,1,0}}      \expandafter\def\csname LT6\endcsname{\color[rgb]{0,0,0}}      \expandafter\def\csname LT7\endcsname{\color[rgb]{1,0.3,0}}      \expandafter\def\csname LT8\endcsname{\color[rgb]{0.5,0.5,0.5}}    \else
      \def\colorrgb#1{\color{black}}      \def\colorgray#1{\color[gray]{#1}}      \expandafter\def\csname LTw\endcsname{\color{white}}      \expandafter\def\csname LTb\endcsname{\color{black}}      \expandafter\def\csname LTa\endcsname{\color{black}}      \expandafter\def\csname LT0\endcsname{\color{black}}      \expandafter\def\csname LT1\endcsname{\color{black}}      \expandafter\def\csname LT2\endcsname{\color{black}}      \expandafter\def\csname LT3\endcsname{\color{black}}      \expandafter\def\csname LT4\endcsname{\color{black}}      \expandafter\def\csname LT5\endcsname{\color{black}}      \expandafter\def\csname LT6\endcsname{\color{black}}      \expandafter\def\csname LT7\endcsname{\color{black}}      \expandafter\def\csname LT8\endcsname{\color{black}}    \fi
  \fi
    \setlength{\unitlength}{0.0500bp}    \ifx\gptboxheight\undefined      \newlength{\gptboxheight}      \newlength{\gptboxwidth}      \newsavebox{\gptboxtext}    \fi    \setlength{\fboxrule}{0.5pt}    \setlength{\fboxsep}{1pt}\begin{picture}(4818.00,2834.00)    \gplgaddtomacro\gplbacktext{      \csname LTb\endcsname      \put(784,512){\makebox(0,0)[r]{\strut{}1e-20}}      \put(784,938){\makebox(0,0)[r]{\strut{}1e-19}}      \put(784,1364){\makebox(0,0)[r]{\strut{}1e-18}}      \put(784,1789){\makebox(0,0)[r]{\strut{}1e-17}}      \put(784,2215){\makebox(0,0)[r]{\strut{}1e-16}}      \put(784,2641){\makebox(0,0)[r]{\strut{}1e-15}}      \put(880,352){\makebox(0,0){\strut{}$0$}}      \put(1336,352){\makebox(0,0){\strut{}$5$}}      \put(1792,352){\makebox(0,0){\strut{}$10$}}      \put(2248,352){\makebox(0,0){\strut{}$15$}}      \put(2705,352){\makebox(0,0){\strut{}$20$}}      \put(3161,352){\makebox(0,0){\strut{}$25$}}      \put(3617,352){\makebox(0,0){\strut{}$30$}}      \put(4073,352){\makebox(0,0){\strut{}$35$}}      \put(4529,352){\makebox(0,0){\strut{}$40$}}    }    \gplgaddtomacro\gplfronttext{      \csname LTb\endcsname      \put(128,1576){\rotatebox{-270}{\makebox(0,0){\strut{}$\Delta E$}}}      \put(2704,112){\makebox(0,0){\strut{}time ($\omega_p^{-1}$)}}      \csname LTb\endcsname      \put(3794,2498){\makebox(0,0)[r]{\strut{}Case A}}      \csname LTb\endcsname      \put(3794,2338){\makebox(0,0)[r]{\strut{}Case B}}      \csname LTb\endcsname      \put(3794,2178){\makebox(0,0)[r]{\strut{}Case C}}      \csname LTb\endcsname      \put(3794,2018){\makebox(0,0)[r]{\strut{}Case D}}    }    \gplbacktext
    \put(0,0){\includegraphics{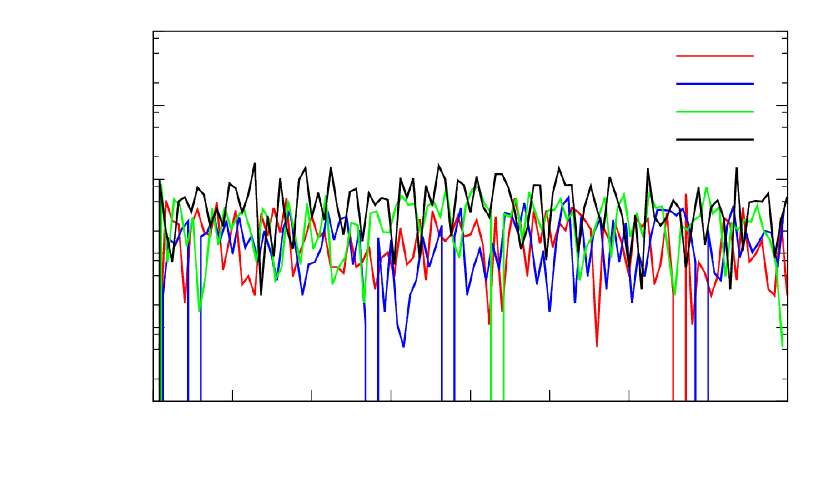}}    \gplfronttext
  \end{picture}\endgroup
   \caption{Variation of the total energy in the system for cases A, B, C, and D.}
  \label{fig-feEnergy}
\end{figure}

We tested the energy conservation of ECsim-CYL for different grid sizes and
time steps and the results are shown in Figure \ref{fig-feEnergy} from the cases in table 1.  Like the ECsim code,
ECsim-CYL conserves the energy down to the numerical accuracy of the solver for all cases. 
In this case, the solver tolerance is set down to machine precision and energy conservation is equal to machine precision.  
For all practical purposes, the energy conservation is exact, as has been proven mathematically, regardless of the dimensionality of the problem \citep{lapenta2016exactly}. 
The cylindrical implementation retains this same property as in the Cartesian case.

The energy conservation has three benefits. First, energy is a key conserved quantity in Nature and it is clearly desirable to preserve the energy 
conservation in numerical algorithms.  Second, non-conservation of energy can lead to numerical instability, a condition where spurious energy can feed erroneous growth of modes that do not exist. 
Imposing energy conservation removes this risk. Third, the energy conservation provides a simple scalar quantity to check the accuracy of the code (i.e. total energy)
and can be utilized as a powerful debugging tool.

\subsection{Divergence Conservation}
\begin{figure}[h]
  \center
  \includegraphics[width=\columnwidth]{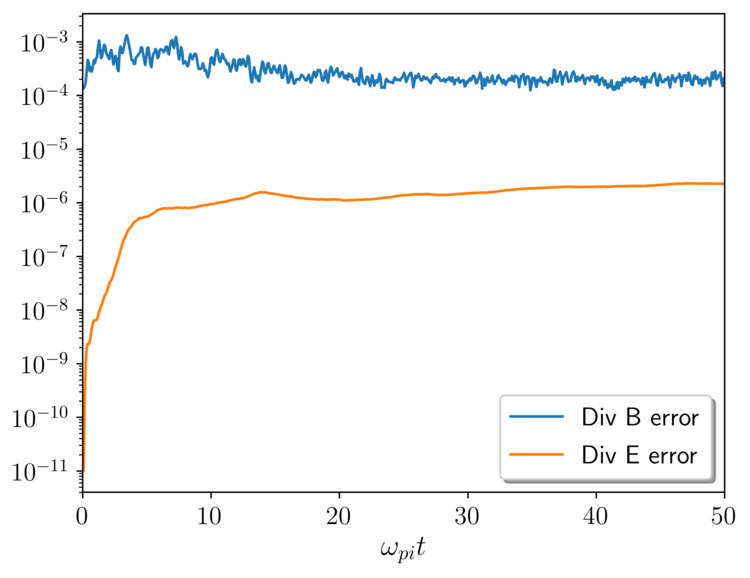}
  \caption{Error in the conservation of the divergence in Maxwell's equations, from simulation run C. The errors are normalized, for $\nabla \cdot \mathbf{B}$ to the current and for  $\nabla \cdot \mathbf{E}$ to the total charge.}
  \label{divergence}
\end{figure}

The choice of finite volume operators reported in Section 2.1 is second order accurate and leads to exact conservation as shown above because the differential operators are integrated over finite volumes and the result is expressed via a surface integral. The conservation properties follow from the fact that the surface integral is identical for both cells sharing the same surface. 
This property comes to a cost: the operators are not mimetic \citep{hyman1999mimetic}, i.e. the known null properties of combined operators, namely the divergence of the curl, are not satisfied exactly. This can lead to numerical errors in code results. 
Figure \ref{divergence} reports the $L_{2}$ norm of the error in the two divergence conditions in Maxwell's equations.

The equation for B (eq.(1)), advances the magnetic field with the curl of the electric field. If the operators were mimetic, the divergence of the magnetic field would always be exactly zero. Since the operators are not mimetic, this leads to the  divergence error shown in Fig. \ref{divergence}, defined as
\begin{equation}
\mathcal{E}_{\nabla \cdot \mathbf{B}} =\frac{\displaystyle \int_{ V_{c}}|\nabla \cdot \mathbf{B}|^{2}dV}{\displaystyle \int_{ V_{c}}|\nabla \times \mathbf{B}|^{2}dV}
\end{equation}
The normalization is typical in the evaluation of the divergence error for B and uses the $L_{2}$ norm of the curl, essentially the current. In comparison, the error in the Gauss theorem is defined as
\begin{equation}
\mathcal{E}_{\nabla \cdot \mathbf{B}} =\frac{\displaystyle \int_{ V_{c}}|(\nabla \cdot \mathbf{E} -\rho|^{2}dV}{\displaystyle \left(\int_{ V_{c}}\rho_{i}dV\right)^{2}},
\end{equation}
with normalization using a total charge of one species in the system (we use positive ions since the total charge including electrons would be zero because of charge neutrality).   
In the present case, both errors saturate quickly in time at a level between  $10^{-4}$ in the case of magnetic field and $10^{-6}$ in the case of electric field.  Considering very weak magnetic fields in the case of spherically symmetric plasma expansion, we find that the errors related to non-mimetic operators are small and would not affect the accuracy of the simulation. 

\subsection{Ambipolar diffusion}

\begin{figure}[h]
  \center
  \includegraphics[width=3.0in]{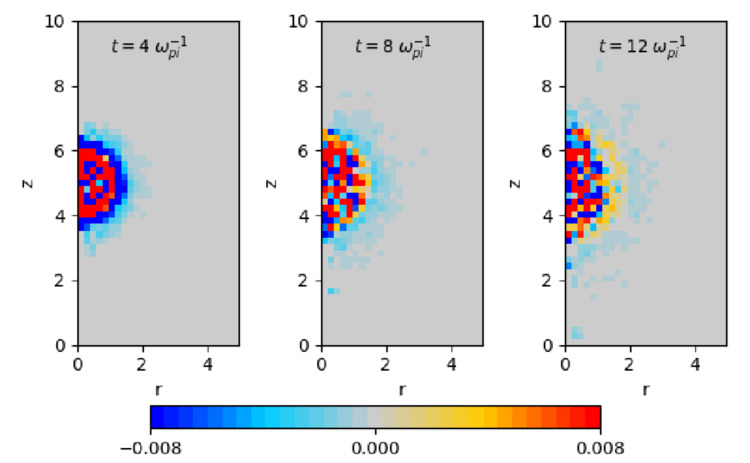}
  \includegraphics[width=3.0in]{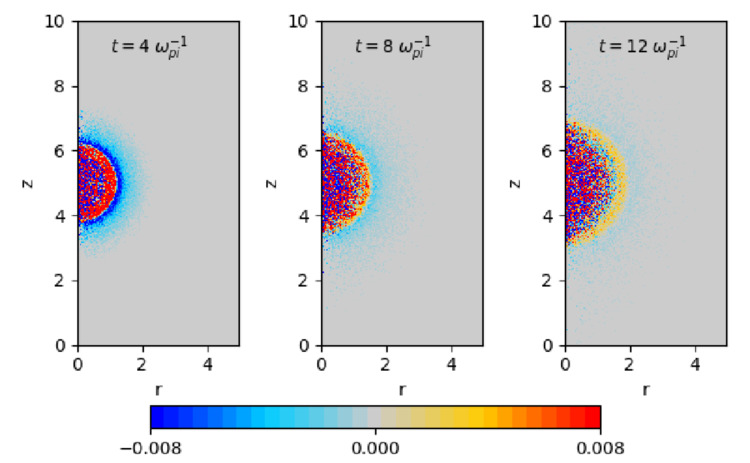}
  \caption{Net charge density profiles at different time steps for two different spatial resolution:  
           $\Delta r =\Delta z = 0.2 \, d_i$ --Case A-- (top panel)
           , and $\Delta x = 0.02 \, d_i$ --Case D-- (bottom panel).}
  \label{fig-feRho}
\end{figure}

As previously discussed, free expansion of a plasma sphere in vacuum will induce charge separation
between ions and electron, self-consistent ambipolar electric fields and change in particle velocity
distribution. \refFig{fig-feRho} shows the net charge density profile for three different times during the
plasma expansion, at $t  = 4 / \omega_{pi}$ , $t  = 8/\omega_{pi}$, and $t  = 12/\omega_{pi}$ for the cases A and D, showing good
agreement between the two cases for all three times.  While the spatial profile of charge
separation is better defined in the high-resolution case of D, the charge separation is still
clearly visible in the case of A where the spatial resolution is 10 times coarser in exchange for
1,000 times faster computation time.

\begin{figure}[h]
  \center
  \includegraphics[width=3.0in]{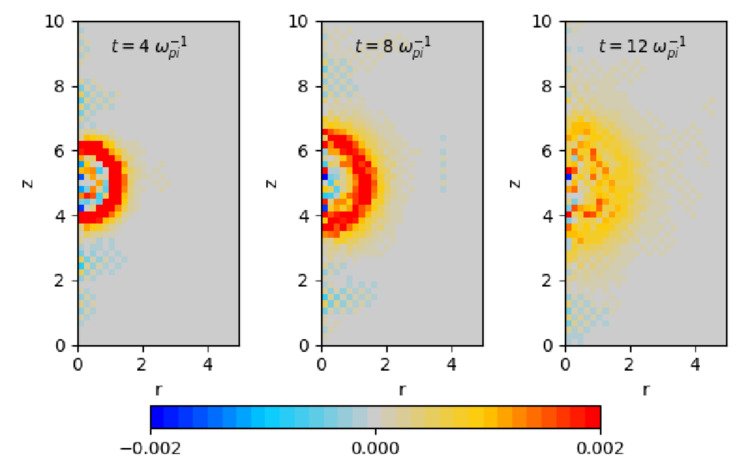}
  \includegraphics[width=3.0in]{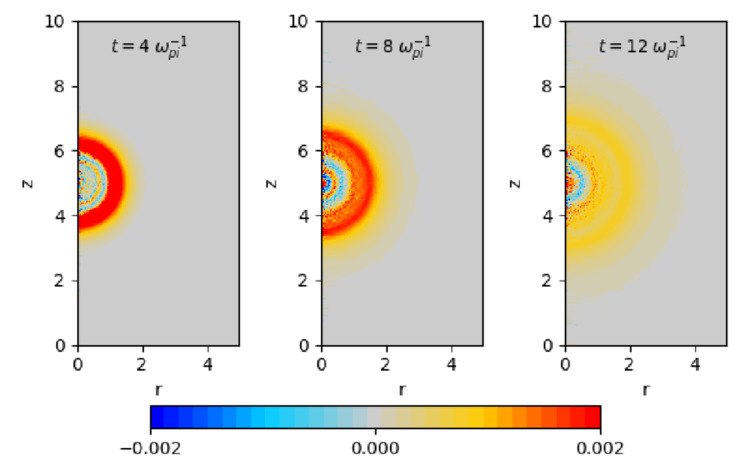}
  \caption{Profiles of electric field in the radial direction (in spherical coordinates) at different times for two different spatial resolutions:  
           $\Delta r =\Delta z = 0.2 \, d_i$ --Case A-- (top panel)
           , and $\Delta x = 0.02 \, d_i$ --Case D-- (bottom panel).}
  \label{fig-feE}
\end{figure}

\refFig{fig-feE} shows the spatial profile of electric field in the radial direction for $t  = 4 /
\omega_{pi}$ , $t  = 8/\omega_{pi}$, and $t  = 12/\omega_{pi}$ for the cases A and D.  Positive values, marked in red, indicate the
electric field pointing radially outward and negative values, marked in blue, are for the radially
inward electric field.  Similar to the net charge density profile shown in \refFig{fig-feRho}, the ECsim-CYL
results show  the ambipolar electric field for both cases of A and D. The comparison is positive. There re of course differences and the details are noisier and less defined in the low resolution run, but one critical aspect is very similar: the extension of the expansion is remarkably similar. Despite a big difference in spatial resolution and computation speed, the low resolution run is still capable of capturing the essential nature of the expansion. Especially remarkable is the correct sign and location of the outermost double layer made of a positive shell surrounded by a negative halo (see \refFig{fig-feRho}). This feature is present in all resolution. Additionally, the radial location of these features is correct even at low resolution.

\begin{figure}[h]
  \center
  \includegraphics[width=2.5in]{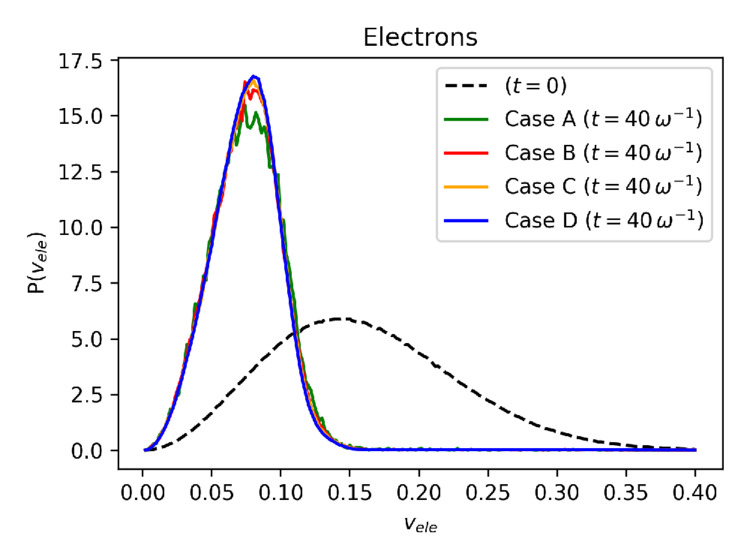}
  \includegraphics[width=2.5in]{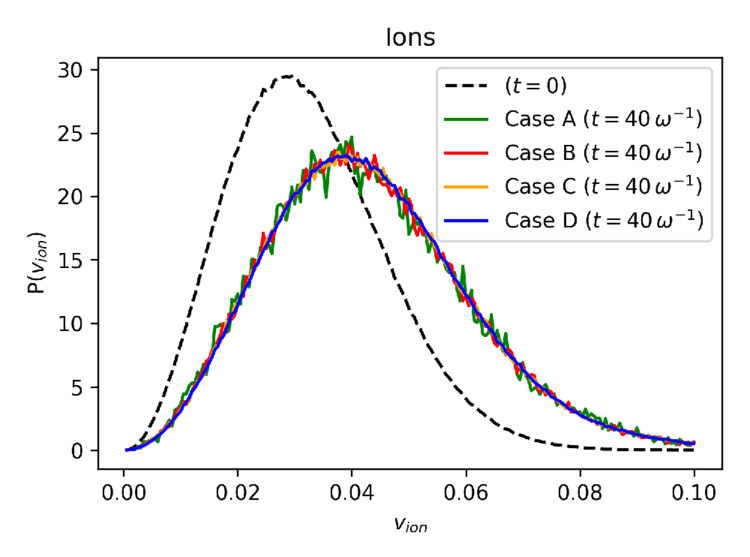}
  \caption{Electron (top panel) and ion (bottom panel) velocity distribution at the beginning of the simulation (dashed lines) and at 
           $t=40 / \omega_{pi}$ (solid lines) from cases A (green), B (red), C (orange) and D (blue).}
  \label{fig-vel}
\end{figure}

\begin{table}[h]
  \center {\small
  \begin{tabular}{|ll|cccc|}
  \hline
            &           & $KE$ & $v_{\text{drift}}(r)/c$ & $v_{\text{drift}}(\theta)/c$ & $v_{\text{drift}}(\phi)/c$ \\
  \hline 
   $t=0$    & electrons & $2.00 \; 10^{-4}$ & N/A & N/A & N/A \\
            & ions      & $2.00 \; 10^{-4}$ & N/A & N/A & N/A \\
  \hline
   A ($t=40/ \omega_{pi}$)  & electrons & $4.24 \; 10^{-5}$ & $3.17 \; 10^{-2}$ & $9.47 \; 10^{-5}$ & $6.89 \; 10^{-4}$ \\
                            & ions      & $3.57 \; 10^{-4}$ & $3.08 \; 10^{-2}$ & $1.76 \; 10^{-4}$ & $2.81 \; 10^{-5}$ \\
  \hline
   B ($t=40/ \omega_{pi}$)  & electrons & $4.39 \; 10^{-5}$ & $3.16 \; 10^{-2}$ & $1.13 \; 10^{-5}$ & $2.84 \; 10^{-4}$ \\
                            & ions      & $3.58 \; 10^{-4}$ & $3.14 \; 10^{-2}$ & $2.46 \; 10^{-4}$ & $8.27 \; 10^{-5}$ \\
  \hline
   C ($t=40/ \omega_{pi}$)  & electrons & $4.17 \; 10^{-5}$ & $3.20 \; 10^{-2}$ & $5.94 \; 10^{-6}$ & $1.98 \; 10^{-5}$ \\
                            & ions      & $3.58 \; 10^{-4}$ & $3.19 \; 10^{-2}$ & $2.38 \; 10^{-5}$ & $1.96 \; 10^{-5}$ \\
  \hline
   D ($t=40/ \omega_{pi}$)  & electrons & $4.13 \; 10^{-5}$ & $3.20 \; 10^{-2}$ & $1.40 \; 10^{-5}$ & $4.28 \; 10^{-5}$ \\
                            & ions      & $3.58 \; 10^{-4}$ & $3.21 \; 10^{-2}$ & $1.28 \; 10^{-5}$ & $4.97 \; 10^{-6}$ \\
  \hline
   A-1836 ($t=0$)     & electrons & $2.71 \; 10^{-6}$ & N/A & N/A & N/A \\
                      & ions      & $2.71 \; 10^{-6}$ & N/A & N/A & N/A \\
  \hline
   A-1836 ($t=343/ \omega_{pi}$)  & electrons & $3.55 \; 10^{-7}$ & $3.33 \; 10^{-3}$ & $6.63 \; 10^{-4}$ & $1.12 \; 10^{-4}$ \\
                                 & ions      & $5.07 \; 10^{-6}$ & $3.49 \; 10^{-3}$ & $2.63 \; 10^{-7}$ & $3.69 \; 10^{-6}$ \\
  \hline
   D-1836 ($t=0$)     & electrons & $2.71 \; 10^{-6}$ & N/A & N/A & N/A \\
                      & ions      & $2.71 \; 10^{-6}$ & N/A & N/A & N/A \\
  \hline
   D-1836 ($t=343/ \omega_{pi}$)  & electrons & $3.40 \; 10^{-7}$ & $3.38 \; 10^{-3}$ & $1.11 \; 10^{-5}$ & $6.15 \; 10^{-5}$ \\
                                 & ions      & $5.08 \; 10^{-6}$ & $3.50 \; 10^{-3}$ & $2.62 \; 10^{-5}$ & $1.27 \; 10^{-6}$ \\
  \hline
  \end{tabular}
  \caption{Kinetic energy (in code units) and drift velocities of electrons and ions during the free expansion.}
  \label{tab-drift}
  }
\end{table}

The role of self-consistent ambipolar electric fields is to balance the rate of expansion for electrons
and ions by transferring energy from lighter electrons to heavier ions, thus modifying their
velocity distribution.  \refFig{fig-vel} shows the velocity distribution of electrons and ions at $t=0$ and $t=40/ \omega_{pi}$ for case A, B,
C and D.  At $t=0$, the electron and ion distribution are given as Maxwellian with the same temperature. 
As the plasma expands, the difference in electron and ion mass and their respective velocity leads to 
the charge separation and the ambipolar electric fields as shown in \refFig{fig-feRho} and
\refFig{fig-feE}.  As shown in \refFig{fig-vel}, the electric field slows down the electrons and shifts its velocity distribution
toward lower velocities. For ions, the electric fields accelerate
ions and shift its velocity distribution toward higher velocities.  It is noted that the good
agreements are observed for all four cases despite their difference in spatial resolution and
computation time.  In addition, the \refTab{tab-drift} shows the total kinetic energy of electron and ions at
$t=40 \; \omega_{pi}$ with their respective drift velocity in $r$, $\theta$, $\phi$ direction for case A, B, C and D,
where $r$, $\theta$, $\phi$ are three orthogonal directions of spherical coordinates.  As expected for the
spherical expansion, the plasma drift velocity is mostly in the radial direction with small
fluctuations in $\theta$ and $\pi$ direction.  Furthermore, the total energy of the system is well
preserved even though significant energy transfer occurs between electrons and ions via ambipolar
electric fields.  Note that small differences in the kinetic energy is due to the difference in
stored energy in electric and magnetic fields in the system, which is 3-4 orders of magnitude
smaller than the kinetic energy in the system. \refTab{tab-drift} further validates that  the ECsim-CYL properly 
describes ambipolar plasma diffusion leading to equalized ion and electron drift velocity over a wide range of spatial
resolution.

\subsection{Ion Mass Ratio}

As additional tests to validate the ECsim-CYL implementation, we have repeated the case A and D with an ion mass
of $m_i = 1836 m_e$ while keeping the electron mass the same.  As shown in \refTab{tab-free}, the case A-1836 uses the
same spatial resolution as the case A, and the case D-1836 uses the same spatial resolution as the case
D.  It is noted that the ion mass of $m_i=1836$ represents hydrogen ions and known as a realistic ion
mass.  Same as the cases A and D, the initial charge density of a homogeneous plasma sphere
for the cases A-1836 and D-1836 is set for the sphere radius to $R_s  = 1 \;d_i$ with the electron thermal
velocity of $v_{\text{th},\text{e}}^r = v_{\text{th},\text{e}}^\theta = v_{\text{th},\text{e}}^z = 0.1   \, c$. Due to the increased ion mass, the ion thermal velocity is 
reduced from $v_{\text{th},\text{i}}^r = v_{\text{th},\text{i}}^\theta= v_{\text{th},\text{i}}^z = 0.02 \, c$ to $0.00233 \, c$ or by a factor of sqrt(1836/25) to keep 
the ion temperature equal to the electron temperature.  The radius and length of the cylindrical simulation 
domain are kept the same when normalized by ion inertial length with R  = 5 di and L  = 10 di.  
Also, we utilized the same CFL condition of $v_{\text{th},\text{e}} \cdot \Delta t/ \Delta x  = 0. 2$.  

\begin{figure}[h]
  \center
  \includegraphics[width=2.5in]{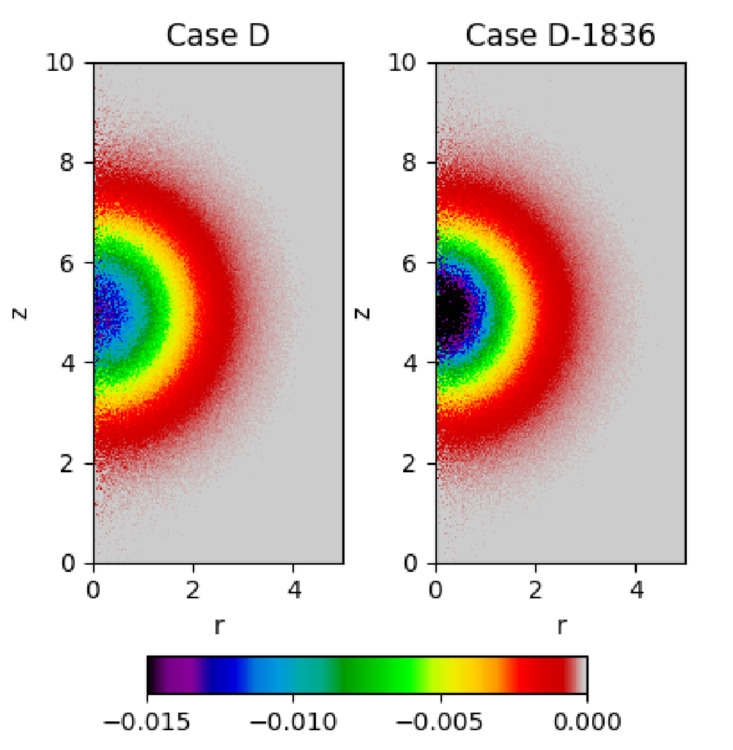}
  \caption{Electron charge density at $t= 40 / \omega_{pi}$ for case D (left) and at $t= 343 / \omega_{pi}$ for case D-1836 (right)}.
  \label{fig-eleM1836}
\end{figure}

Here we note two main changes in the simulation related to the change in ion mass.  The first is
the reduced ion velocity for the same plasma temperature.  Heavier and slower ions will require
longer time to expand outward.  While the ambipolar electric fields will still accelerate ions by
transferring energy from electrons, the outward ion drift velocity will be reduced by $\sqrt{1836/25}$ 
or $8.57$ compared to the ion mass of $m_i=25 m_e$.  This corresponds to the increase in simulation cycles by
$\approx 8.57$ times for the plasma to reach the same radial extension.  \refFig{fig-eleM1836}
compares the electron charge density profile for the case D at $t=40 / \omega{pi}$ and for the case D-1836 at
$t= 343 / \omega{pi}$, showing a slower rate of plasma expansion for heavier ions.  In comparison, the time step
for the cases A-1836 and D-1836 is determined by the CFL condition, thus there is no need to adjust the time
step as long as the same value of $v_{\text{th},\text{e}}$ and $\Delta x$ are used.  The second change related to the ion mass
is the ratio between ion inertial length and electron Debye length.  For a fixed ion inertial
length, the Debye length will decrease by a square root of ion mass, corresponding to an initial
plasma sphere radius of $R \approx 430 \, \lambda_e$ for the cases of A-1836 and D-1836 instead of $R = 50 \, \lambda_e$ for 
the cases of A and D.

\begin{figure}[h]
  \center
  \includegraphics[width=2.5in]{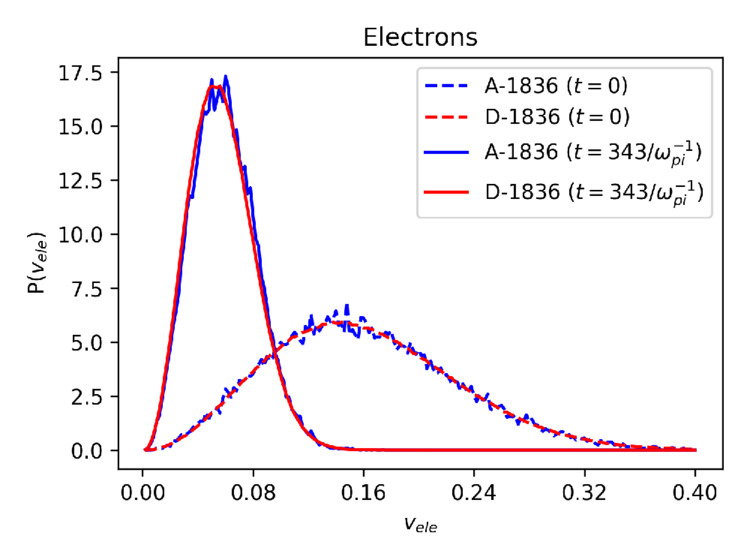}
  \includegraphics[width=2.5in]{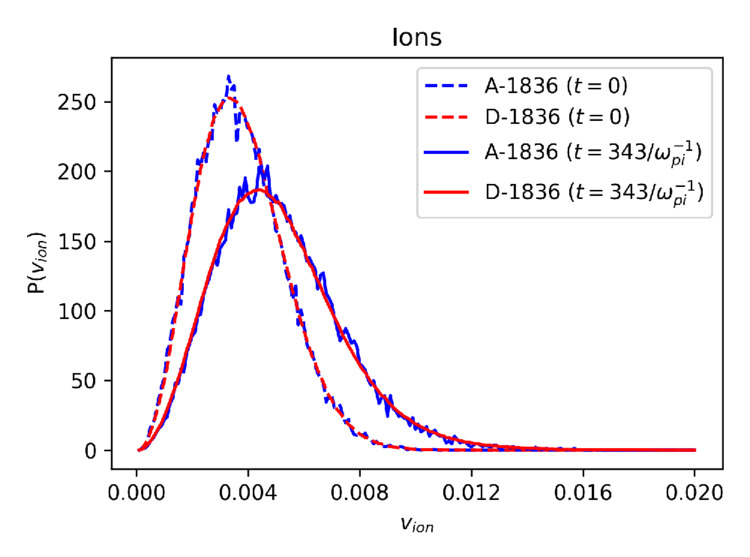}
  \caption{Electron (top panel) and ion (bottom panel) velocity distribution at the beginning of the simulation (dashed lines) and at $t= 343 / \omega_{pi}$
           (solid lines) for cases A-1836 (blue) and D-1836 (red). All particles in the system are included in the analysis.}
  \label{fig-vel_m}
\end{figure}

\refFig{fig-vel_m} shows the velocity distribution of electrons and ions at $t=0$ and $t=343/\omega_{pi}$ 
for the cases A-1836 and D-1836. As in \refFig{fig-vel}, a good agreement between
these two cases is observed for the electron and the ion velocity distribution despite their difference in
spatial resolution and computation time.  It is noted that with the increased in ion mass, the peak
of electron velocity distribution occurs for $0.055 \; c$ at $t=343/\omega_{pi}$ from $0.137 \;c$ at $t=0$.  
In comparison, the peak of electron velocity distribution shifts to $0.081 \;c$ at $t= 40/\omega_{pi}$ from $0.137 \;c$ 
at $t=0$ in the case of $m_i = 25 \; m_e$, as shown in \refFig{fig-vel}. As for ions, the peak of ion velocity distribution
shifts to $0.0044 \;c$ at $t=343/\omega_{pi}$ from $0.0033 \;c$ at $t=0$ in the cases $m_i=1836$, while the peak shift from $0.028\; c$
to $0.038 \; c$ in the cases of $m_i=25\; m_e$.  In addition, the \refTab{tab-drift} shows the total kinetic energy of ions and
electrons at $t=343/\omega_{pi}$ with their respective drift velocity in $r$, $\theta$, $\phi$ direction for cases A-1836 and D-1836.  
These results indicate that more energy is transferred from electrons to ions via
ambipolar electric fields with increased ion mass. The results from the \refFig{fig-eleM1836}, \refFig{fig-vel_m} and the \refTab{tab-drift}
shows ECsim-CYL properly simulates ambipolar plasma expansion for a realistic ion mass of $m_i=1836 \;m_e$
while using the same time step and spatial resolution of reduced ion mass cases of $m_i=25 \; m_e$. 

\subsection{Grid resolution with respect to Debye length}

\begin{figure}[h]
  \center
  \includegraphics[width=4.8in]{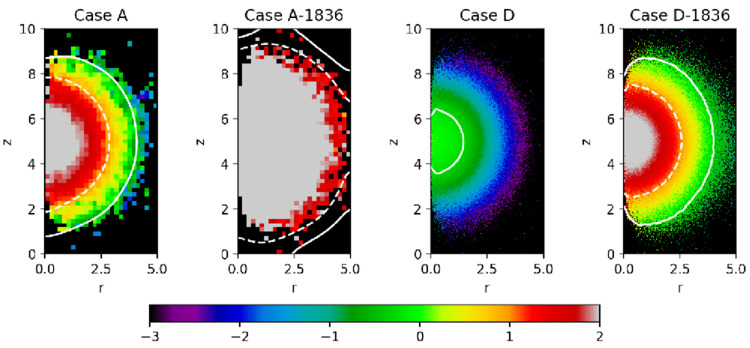}
  \caption{Debye length (in code units with logarithmic scale) at $t=40 / \omega_{pi}$ for cases A and D and at
$t=343 / \omega_{pi}$ for cases A-1836 and D-1836. The solid white line indicates the isocontour of $\Delta x  =
\lambda_D / 2$ and the dashed white line indicates the isocontour of $\Delta x  = 10 \;\lambda_D$.}
  \label{fig-Debye}
\end{figure}

One of the noteworthy findings from the free expansion test is the robustness of ECsim-CYL as it
accurately simulates the ambipolar diffusion even for a very coarse spatial resolution.
\refFig{fig-Debye} shows the ratio of spatial resolution ($\Delta x$) and electron Debye length for 
cases A and D at $t= 40/\omega_{pi}$ and cases A-1836 and D-1836 at $t= 343/\omega_{pi}$ in logarithmic scale with 
the solid white line indicating the isocontour of $\Delta x  = \lambda_D / 2$. In the case of D, the 
spatial resolution is sufficient to resolve the electron Debye length except for the
core region inside the white line.  On the other hand, the electron Debye length is not resolved in
most of the simulation domain in the case of A and D-1836.  The spatial resolution is even coarser in the case
of A-1836 where the spatial resolution is more than 10 times the electron Debye length in most of the
system.  The results in \refFig{fig-Debye} indicate that the ECsim-CYL can provide the accurate description
of ambipolar effects even when the spatial resolution is many times larger than the Debye length.  

Since the ambipolar diffusion is one of the critical conditions that must be satisfied in simulation
to properly describe the plasma motion, this result offers a tantalizing prospect of a full fusion
device simulation using ECsim-CYL. Typical plasma parameters for a fusion reactor are the
following: a reactor size of $\approx 5$ meter in linear dimension, a plasma density of $1\times10^{14}$ 
cm and a plasma temperature of $20$ keV leading to an electron Debye length of $0.01$ cm. This large difference
between the system size and the Debye length is one of the major challenges in simulation of a
full-scale fusion device if one needs to resolve the electron Debye length to include accurate description of 
ambipolar diffusion in the system.  For example, in a fully three dimensional reactor
configuration without a symmetric axis such as tokamaks and stellarators, this requirement translates
to a very large number of cells on the order of 100,000 x 100,000 x 100,000 cells or $10^{15}$ cells.
In comparison, if the system possesses axisymmetry and is free from plasma instability that can
break the symmetry, one can utilize a reduced dimension code such as ECsim-CYL to greatly reduce the
required number of cells.  Additional reduction in a required cell number may be achieved by taking
an advantage of ECsim-CYL which can properly describe the critical ambipolar diffusion of the plasma
system with the use of spatial resolution much larger than the electron Debye length.  In the case
where spatial resolution of 10 Debye lengths is adequate, the ECsim-CYL can simulate an entire
fusion reactor with the cell size of 5,000 x 5,000 or $2.5\times10^{7}$ cells which is well within the
numerical capability of modern high performance computing facilities.  Another advantage of ECsim-CYL
is that the size of time step is controlled by the CFL condition.  Therefore, a bigger time step can be
used for a larger grid size, which is desirable for a large system domain since the particle travel
time across the system is significant.
 \section{Conclusions}

We have presented ECsim-CYL, a PIC code in cylindrical coordinates for axially symmetric simulations 
based on the previously developed ECsim code.  We have discussed the discretization used to numerically solve
the field equations in cylindrical coordinates and the strategy for moving the particles. We have studied the accuracy
of the particle mover and the field discretization, showing that both the field solver and the particle mover are second- order 
accurate. In section 4 and 5, we have tested the ECsim-CYL code using two well-known problems in plasma physics:
the plasma wave spectrum in an homogeneous plasma inside a cylindrical waveguide and the homogeneous expansion of a plasma sphere in vacuum. 
In the first case, the results given by the code show a good agreement with the analytical formula. While the second case, free expansion of a plasma sphere, 
is more complex and without direct analytical comparison, it provides an excellent example to asses the practical advantages of ECsim-CYL.  
The test results show that the ECsim-CYL implementation accurately preserves the spherical symmetry, conserves the total energy of the system, 
and accurately describes the ambipolar effects over a wide range of spatio-temporal resolution.
In particular, we have shown that the ECsim-CYL properly simulates the critical energy transfer between electrons and ions via self-consistent electric fields
behind the ambipolar diffusion even if the Debye length is poorly resolved with the grid spacing exceeding 10 times the Debye length 
in the simulation domain, a key point of semi-implicit approach.

The  new ECsim-CYL code thus opens exciting opportunities for future research in plasma simulation. 
Using an axial symmetry, the computational time is dramatically reduced when compared with the 3D Cartesian description.
Combined with the relaxed resolution constraint in grid spacing and time step of semi-implicit code, 
this feature opens the door to global simulations of laboratory experiments and natural plasmas in a fully kinetic approach.
Specifically, the ECsim-CYL may open the door to global simulations of fusion and astrophysical plasma systems 
where the gap in scales between the Debye length and the system size is beyond the reach of explicit methods even for next generation exascale supercomputers.
Provided that the plasma systems possess an axial symmetry, it becomes possible to conduct full device simulations using ECsim-CYL 
at the fully kinetic level where electrons and ions are both described as particles.

\section*{Acknowledgments}
The research leading to these results has received funding from the European Community's Horizon 2020 (H2020)
Funding Program under Grant Agreement n. 754304 (Project DEEP-EST) and has been supported by R\&D Agreement
between Energy Matter Conversation Corporation (EMC2) and KU Leuven R\&D (Contract \# 2017/771). The computations
were carried out at the Flemish Supercomputer Centre (VSC). This research used resources of the National
Energy Research Scientific Computing Center, a DOE Office of the Science User Facility supported by the Office of
Science of the U.S. Department of Energy under Contract No. DE-AC02-05CH11231. Authors would like to thank
Dr. Nicholas Krall from Energy Matter Conversion Corporation for his insightful discussion regarding ECsim-CYL results.

\linenumbers


\begin{thebibliography}{10}
\expandafter\ifx\csname url\endcsname\relax
  \def\url#1{\texttt{#1}}\fi
\expandafter\ifx\csname urlprefix\endcsname\relax\def\urlprefix{URL }\fi
\expandafter\ifx\csname href\endcsname\relax
  \def\href#1#2{#2} \def\path#1{#1}\fi

\bibitem{birdsall-langdon}
C.~Birdsall, A.~Langdon, Plasma Physics Via Computer Simulation, Taylor \&
  Francis, London, 2004.

\bibitem{hockney-eastwood}
R.~Hockney, J.~Eastwood, Computer simulation using particles, Taylor \&
  Francis, 1988.

\bibitem{markidis2011energy}
S.~Markidis, G.~Lapenta, The energy conserving particle-in-cell method, J.
  Comput. Phys. 230~(18) (2011) 7037--7052.

\bibitem{Chen-jcp-2011}
G.~Chen, L.~Chac\'on, D.~Barnes, An energy- and charge-conserving, implicit,
  electrostatic particle-in-cell algorithm, J. Comput. Phys. 230 (2011) 7018.

\bibitem{morse1971numerical}
R.~Morse, C.~Nielson, Numerical simulation of the weibel instability in one and
  two dimensions, The Physics of Fluids 14~(4) (1971) 830--840.

\bibitem{lapenta2012particle}
G.~Lapenta, Particle simulations of space weather, J. Comput. Phys. 231~(3)
  (2012) 795--821.

\bibitem{brackbill-forslund}
J.~Brackbill, D.~Forslund, An implicit method for electromagnetic plasma
  simulation in two dimension, J. Comput. Phys. 46 (1982) 271.

\bibitem{Brackbill:1985}
J.~U. {Brackbill}, B.~I. {Cohen} (Eds.), {Multiple time scales.}, 1985.

\bibitem{chen2014fluid}
G.~Chen, L.~Chac{\'o}n, C.~A. Leibs, D.~A. Knoll, W.~Taitano, Fluid
  preconditioning for newton--krylov-based, fully implicit, electrostatic
  particle-in-cell simulations, J. Comput. Phys. 258 (2014) 555--567.

\bibitem{directimplicit}
A.~Langdon, B.~Cohen, A.~Friedman, Direct implicit large time-step particle
  simulation of plasmas, J. Comput. Phys. 51 (1983) 107--138.

\bibitem{welch2004implementation}
D.~R. Welch, D.~V. Rose, R.~E. Clark, T.~C. Genoni, T.~Hughes, Implementation
  of an non-iterative implicit electromagnetic field solver for dense plasma
  simulation, Computer physics communications 164~(1) (2004) 183--188.

\bibitem{ipic3d}
S.~Markidis, G.~Lapenta, Rizwan-uddin, Multi-scale simulations of plasma with
  {iPIC3D}, Mathematics and Computers and Simulation 80 (2010) 1509--1519.

\bibitem{lapenta2016exactly}
G.~Lapenta, Exactly energy conserving semi-implicit particle in cell
  formulation, J. Comput. Phys. 334 (2017) 349--366.

\bibitem{Lapenta2017}
{G. Lapenta, D. Gonzalez-Herrero and E. Boella}, Multiple-scale kinetic
  simulations with the energy conserving semi-implicit particle in cell method,
  J. Plasma Phys. 83 (2017) 705830205.

\bibitem{GonzalezHerrero2018}
{D. Gonzalez-Herrero, E. Boella and G. Lapenta}, Performance analysis and
  implementation details of the energy conserving semi-implicit method code
  (ecsim), Comput. Phys. Comm. (2018) doi.org/10.1016/j.cpc.2018.03.020.

\bibitem{burgess1992mass}
D.~Burgess, D.~Sulsky, J.~Brackbill, Mass matrix formulation of the flip
  particle-in-cell method, J. Comput. Phys. 103~(1) (1992) 1--15.

\bibitem{Ringle2011}
{R. Ringle}, 3dcylpic—a 3d particle-in-cell code in cylindrical coordinates
  for space charge simulations of ion trap and ion transport devices, J. Mass
  Spectrom 303 (2011) 42--50.

\bibitem{Wallace1986}
{J.M. Wallace, J.U. Brackbill and D.W. Forslund}, An implicit moment
  electromagnetic plasma simulation in cylindrical coordinates, Comput. Phys.
  Comm. 63 (1968) 434--457.

\bibitem{lapenta2011democritus}
G.~Lapenta, Democritus: An adaptive particle in cell (pic) code for
  object-plasma interactions, Journal of Computational Physics 230~(12) (2011)
  4679--4695.

\bibitem{Delzanno2013}
{G.L. Delzanno, E. Camporeale, J.D. Moulton, J.E. Borovsky, E.A. MacDonald, and
  M.F. Thomsen}, Cpic: A curvilinear particle-in-cell code for
  plasma–material interaction studies, IEEE Trans. Plasma Sci. 41 (2013)
  3577--3587.

\bibitem{Chacon2016}
{L. Chacon, G. Chen}, A curvilinear, fully implicit, conservative
  electromagnetic pic algorithm in multiple dimensions 316 (2016) 578--597.

\bibitem{leveque2}
E.~A.~D. R.~J.~LeVeque, D.~Mihalas, E.~Muller, Computational Methods for
  Astrophysical Fluid Flow, Springer, 1998.

\bibitem{petsc-web-page}
S.~Balay, S.~Abhyankar, M.~F. Adams, J.~Brown, P.~Brune, K.~Buschelman,
  L.~Dalcin, V.~Eijkhout, W.~D. Gropp, D.~Kaushik, M.~G. Knepley, L.~C.
  McInnes, K.~Rupp, B.~F. Smith, S.~Zampini, H.~Zhang, H.~Zhang,
  \href{http://www.mcs.anl.gov/petsc}{{PETS}c {W}eb page},
  {http://www.mcs.anl.gov/petsc} (2016).
\newline\urlprefix\url{http://www.mcs.anl.gov/petsc}

\bibitem{riva2017methodology}
F.~Riva, C.~F. Beadle, P.~Ricci, A methodology for the rigorous verification of
  particle-in-cell simulations, Physics of Plasmas 24~(5) (2017) 055703.

\bibitem{Khalil2014}
{Sh. M. Khalil and N.M. Mousa}, Dispersion characteristics of plasma-filled
  cylindrical waveguide, J Theor Appl Phys 8:111 (2014) 14.

\bibitem{KrallTrivelpiece}
N.~A. Krall, A.~W. Trivelpiece, Principles of Plasma Physics, McGraw-Hill,
  1973.

\bibitem{hyman1999mimetic}
J.~M. Hyman, M.~Shashkov, Mimetic discretizations for maxwell's equations,
  Journal of Computational Physics 151~(2) (1999) 881--909.

\end{thebibliography}

\end{document}